\documentclass[fleqn,usenatbib]{mnras}

\usepackage[T1]{fontenc}
\usepackage{ae,aecompl}

\usepackage{graphicx}	
\usepackage{amsmath}	
\usepackage{amssymb}	
\usepackage{bm}
\usepackage{adjustbox}

\usepackage{hyperref}

\def\prl{Phys.~Rev.~Lett.}

\def\jopp{J.~Plasma Phys.}

\def\apjl{ApjL}
\def\pop{Phys.~Plasmas}

\newcommand{\D}[2]{\frac{{\rm d} #2}{{\rm d} #1}}
\newcommand\bcdot{\,\bb{\cdot}\,}
\newcommand\btimes{\,\bb{\times}\,}
\newcommand\bs[1]{\boldsymbol{#1}}
\newcommand\bb[1]{\mbox{\boldmath{$#1$}}}
\newcommand\grad{\bb{\nabla}}
\newcommand{\eb}{\hat{\bb{b}}}
\newcommand{\rmd}{{\rm d}}

\title[CRs in large-amplitude turbulence]{Cosmic ray transport in large-amplitude turbulence with small-scale field reversals}

\author[Kempski et al.]{Philipp Kempski$^{1}$\thanks{E-mail: pkempski@princeton.edu},
Drummond B. Fielding$^{2}$,
Eliot Quataert$^{1}$,
Alisa K. Galishnikova$^{1}$,
\newauthor
Matthew W. Kunz$^{1,3}$,
Alexander A. Philippov$^{4}$,
Bart Ripperda$^{5,6,2}$ 
\\
$^{1}$Department of Astrophysical Sciences, Princeton University, Princeton, NJ 08544, USA \\
$^{2}$Center for Computational Astrophysics, Flatiron Institute, 162 5th Ave, New York, NY 10010, USA\\
$^{3}$Princeton Plasma Physics Laboratory, PO Box 451, Princeton, NJ 08543, USA \\
$^{4}$Department of Physics, University of Maryland, College Park, MD 20742, USA \\
$^{5}$School of Natural Sciences, Institute for Advanced Study, 1 Einstein Drive, Princeton, NJ 08540, USA \\
$^{6}$NASA Hubble Fellowship Program, Einstein Fellow \\
}

\begin{document}
\label{firstpage}
\pagerange{\pageref{firstpage}--\pageref{lastpage}}
\maketitle

\begin{abstract}
The nature of cosmic ray (CR) transport in the Milky Way remains elusive. The predictions of current micro-physical CR transport models in magneto-hydrodynamic (MHD) turbulence are drastically different from what is observed. These models usually focus on MHD turbulence with a strong guide field and ignore the impact of turbulent intermittency on particle propagation. This motivates our studying the alternative regime of large-amplitude turbulence with $\delta B/B_0 \gg 1$, in which intermittent small-scale magnetic field reversals are ubiquitous. We study particle transport in such turbulence by integrating trajectories in stationary snapshots. To quantify spatial diffusion, we use a setup with continuous particle injection and escape, which we term the turbulent leaky box. We find that particle transport is very different from the strong-guide-field case. Low-energy particles are better confined than high-energy particles, despite less efficient pitch-angle isotropization at small energies. In the limit of weak guide field, energy-dependent confinement is driven by the energy-dependent (in)ability to follow reversing magnetic field lines exactly and by the scattering in regions of ``resonant curvature", where the field line bends on a scale that is of order the local particle gyro-radius. We derive a heuristic model of particle transport in magnetic folds that approximately reproduces the energy dependence of transport found numerically.  We speculate that CR propagation in the Galaxy is regulated by the intermittent field reversals highlighted here and discuss the implications of our findings for CR transport in the Milky Way.  
\end{abstract}


\vspace{-55pt}

\section{Introduction}

The lifetime of relativistic cosmic rays (CRs) in the Galaxy is much longer than the light-crossing time of the Milky Way. In the standard paradigm of CR transport, the long residence time is thought to be primarily due to resonant interactions with small-amplitude magnetic-field fluctuations. Small-angle scattering events result in slow pitch-angle diffusion, which leads to the spatial diffusion of CRs along the local magnetic-field direction (\citealt{kp69}; \citealt{skilling71}; \citealt{skilling_1975}; \citealt{zweibel17}).

The origin of the fluctuations that scatter CRs remains unclear. One possibility is that CRs are scattered by magnetic-field fluctuations that are part of an ambient turbulent cascade in the interstellar medium (ISM). Turbulence involves waves propagating in all directions and is thought to result in CR transport that is diffusive (\citealt{skilling_1975}). Alternatively, CRs may themselves excite the waves that scatter them through the CR streaming instability (CRSI; \citealt{kp69}). In this case, CRs excite Alfv\'en waves if their bulk velocity exceeds the Alfv\'en speed. The excited waves propagate down the CR pressure gradient and pitch-angle scatter CRs in the frame of the wave. According to this picture, the streaming instability results in transport that is a combination of streaming and a correction term that is not truly diffusive in the usual sense (\citealt{skilling71};  \citealt{wiener2013}; \citealt{kq2022}). The correction term comes from the damping of Alfv\'en waves excited by the CRSI (\citealt{kc71}; \citealt{lee_volk_1973}; \citealt{farmer_goldreich}; \citealt{wiener18}; \citealt{squire_dust_2021}; \citealt{xu_lazarian_damping}) and increases with increasing damping rate. Notably, the damping prevents the excitation of longer-wavelength modes by less numerous higher-energy CRs ($E\gtrsim100$GeV). As a result, higher-energy CRs are unable to self-confine via the CRSI and an external source of scattering, such as turbulence, is required for confining them. The transport of energetically important GeV CRs may in principle be regulated either by turbulence or by waves excited through the CRSI. In the former case, CR transport is diffusive, while in the latter case the transport is to leading order described by streaming at the local Alfv\'en speed. The nature of how CRs are transported (via streaming or diffusion) therefore depends on the origin of the fluctuations that scatter them. 

The fact that the two origins of fluctuations that scatter CRs result in fundamentally different CR transport models presents an obstacle to formulating predictive models of galaxy evolution. In particular, models in which CRs have a self-limiting streaming velocity, as opposed to those in which CRs diffuse relative to the thermal gas, generally predict very different outcomes in a variety of astrophysical contexts, including  mass outflow in galactic winds or the heating of diffuse gas in galactic halos (e.g., \citealt{guo08}; \citealt{ruszkowski17}; \citealt{wiener_wind}; \citealt{butsky_2018}; \citealt{farber18}; \citealt{hopkins2020_whatabout}; \citealt{qtj_2021_streaming}; \citealt{qtj_2021_diff}).

Observations of CRs in the Milky Way and in other galaxies provide important constraints on the physics of CR transport. Detailed CR energy spectra measured close to Earth are particularly informative as they show a clear energy dependence in CR propagation (\citealt{stone_voyager}; \citealt{aguilar_2015}; \citealt{aguilar_bc}; \citealt{cummings_2016}). In particular, the CR escape time from the Galaxy seems to be approximately consistent with $\tau_{\rm esc} \propto E^{-0.5}$, where $E$ is the CR energy. The  local Milky Way CR spectra can be explained reasonably well using propagation models with a prescribed energy-dependent diffusion coefficient or using simplified microphysical models, e.g., assuming scattering by gyro-resonances and the presence of turbulent magnetic-field fluctuations that obey an isotropic Kolmogorov-like (with power spectrum $P(k) \propto k^{-5/3}$, where $k$ is the wavenumber) or Kraichnan-like ($P(k) \propto k^{-3/2}$) scaling (e.g., \citealt{trotta_2011}; \citealt{blasi12}; \citealt{gaggero_2014}; \citealt{evoli_2018};  \citealt{werhahn_2021}; \citealt{hopkins_cr_pheno}). These simplified approaches are not, however, well motivated by current models of MHD turbulence (\citealt{kq2022}; \citealt{hopkins_sc_et_problems}). In particular, standard quasi-linear theories of CR transport in incompressible MHD turbulence with a strong mean magnetic field  predict negligible confinement of low- and intermediate-energy CRs that have gyro-radii much smaller than the turbulence driving scale (\citealt{chandran_scattering}; \citealt{yan_lazarian_2004}; \citealt{fornieri_2021}). By contrast, an isotropic Kolmogorov-like cascade of magnetic field fluctuations ($\delta B(k) \propto k^{-1/3}$) results in better confinement at lower energies according to quasi-linear predictions.\footnote{For an isotropic  spectrum of magnetic-field fluctuations that satisfy $\delta B(k) \propto k^{-\alpha}$, quasi-linear theory predicts a gyro-resonant scattering frequency $\nu(r_L) \propto r_L^{2 \alpha -1}$, where $r_L$ is the gyro-radius.} The difference between the two turbulence models comes from the fact that the turbulent cascades of MHD Alfv\'en and slow magnetosonic waves produce eddies that are highly elongated along the local magnetic field (\citealt{higdon_1984}; \citealt{gs95}), which drastically suppresses CR scattering (\citealt{chandran_scattering}; \citealt{yan_lazarian_2004}), while eddies in hydrodynamic Kolmogorov turbulence are isotropic. For this reason, \cite{yan_lazarian_2004} proposed that CRs are scattered by the compressible MHD fast-mode cascade, assuming that it is isotropic and obeys a weak-turbulence $P(k) \propto k^{-3/2}$ power spectrum based on the theory of \cite{zakharov_sagdeev_1970} and the simulations analysed by \cite{cho_lazarian}. However, fast modes are subject to significant damping in the volume-filling hot and low-density phases of the ISM and their weak-turbulence cascade is likely to be suppressed by wave steepening (\citealt{kadomtsev1973acoustic}; \citealt{kq2022}). As a result, CR scattering by MHD fast modes may be unimportant, particularly for lower energy CRs.  This implies that there is currently no model of efficient CR scattering in MHD turbulence that is consistent with Milky Way observations.

One plausible remedy is that modern state-of-the-art turbulence simulations highlight the prevalence of intermittency in MHD turbulence (e.g., \citealt{mallet_2015_cb}; \citealt{dong_2018}; \citealt{dong_2022}; \citealt{fielding_plasmoid}), which has not been taken into account in the context of CR transport modelling. Scattering by rare (i.e., not volume filling) but intense (order-unity changes) structures, such as small-scale magnetic-field reversals or plasmoids produced in current sheets (e.g., \citealt{loureiro_boldyrev_2016}; \citealt{mallet_2017_tearing}; \citealt{dong_2022}; \citealt{fielding_plasmoid}) may be crucial for bridging the gap between observations and theories of particle propagation in turbulence. Interestingly, the possible importance of small-scale magnetic folds generated in intermittent turbulence has also been highlighted outside of the CR transport community, e.g., to explain pulsar scintillation observations (\citealt{pen_levin_2014}) or sources of extreme radio wave scattering (\citealt{goldreich_sridhar_folds}). 

The goal of this work is to identify new aspects of particle transport that are not part of existing theories of CR propagation, which usually assume small-amplitude fluctuations and employ quasi-linear theory. We do so by integrating particle trajectories in snapshots of high-resolution MHD turbulence simulations. To study the impact of small-scale bends and reversals in the magnetic-field direction, we focus on large-amplitude turbulence with a weak guide field, i.e.~$\delta B/B_0 \gg 1$. This regime of turbulence produces frequent order-unity variations in the magnetic field, very different from the quasi-linear assumption.  It has received significantly less attention in the CR-transport literature than the strong-guide-field case (see, however, \citealt{subedi_2017} for a discussion of CR transport in synthetic turbulence without a guide field), even though large-amplitude turbulence is likely present in at least parts of the Galaxy. By stretching and folding the magnetic-field lines, large-amplitude turbulence produces field morphologies characterised by small perpendicular reversal scales and regions of high field-line curvature that are somewhat reminiscent of the MHD dynamo (e.g., \citealt{schekochihin_2004}; \citealt{rincon_dynamo}). 

We find that particle transport in the limit of weak guide field is qualitatively very different from the transport in the presence of a strong guide field. Even particles with gyro-radii much smaller than the turbulence injection scale are not fully magnetized\footnote{We consider a particle magnetized if magnetic-field variations on gyro-radius scales are small, i.e.~$r_L^{-1} \gg |\nabla \ln B|$.} and their ability to follow the magnetic folds created by the turbulence is significantly inhibited. In particular, particles that are magnetized in most of the volume can become unmagnetized and escape a field line in regions where the field changes direction rapidly. This inability to follow field lines exactly, alongside pitch-angle scattering by large field-line curvature at the locations where the folds bend, appears to be the dominant factor setting the spatial diffusion of particles (rather than pitch-angle diffusion by low-amplitude waves). These findings suggest that the transport of particles in the presence of magnetic fields organised into fold-like structures, which are a natural outcome of $\delta B/B_0 \gg 1$ turbulence, is fundamentally different from the standard paradigm of CR  propagation and may be important for understanding CR observations in the Galaxy. While our simulations do not probe a sufficiently large range of scales to resolve the disruption of current sheets and the formation of plasmoids, which would contribute an additional source of small-scale bends in the magnetic field, many aspects of the theory considered in this work may also apply to CR transport in plasmoid-like structures.

The paper is structured as follows. We describe our methods for simulating turbulence and integrating particle trajectories in Section~\ref{sec:method}. Section~\ref{sec:method} also introduces the turbulent-leaky-box technique for quantifying CR diffusion. We present results from the turbulent-leaky-box simulations in Section~\ref{sec:leaky_box_results} and provide a model for interpreting these results in Section~\ref{sec:model}. We discuss the importance of our findings for CR transport in the Galaxy in Section~\ref{sec:discussion} and summarise our results in Section~\ref{sec:conclusions}.

Near the completion of this work, we became aware
of a similar effort to study the impact of turbulent intermittency and field-line curvature on the transport of galactic CRs by \cite{lemoine_2023}.

\section{Turbulence simulations and particle transport} \label{sec:method}
We study CR transport in MHD turbulence by integrating particle orbits through stationary snapshots of high-resolution simulations of large-amplitude MHD turbulence. Ignoring the time dependence of the turbulence is well motivated in this particular application, as the CR propagation speeds ($\approx$ speed of light) are much larger than the speeds of the turbulent eddies (subsonic in our simulations). This would not be appropriate, for example, for the GeV CRs if their transport is set by streaming at the Alfv\'en speed, in which case one has to include the time dependence of the large-scale turbulent fields (e.g., \citealt{sampson_2023}). We describe our turbulence simulations and particle-integration methods in more detail below. We also describe our method for measuring the particles' effective transport rate, to which we refer as the `turbulent leaky box'.

\subsection{Turbulence simulations}
For our simulations with a net magnetic flux, we use {\tt Athena++} (\citealt{stone_athena_2008}) to drive isothermal ideal MHD turbulence in a cubic box of size $L=1$ and resolution $1024^3$ with random forcing that follows an Ornstein--Uhlenbeck process (\citealt{uhlenbeck_ornstein_1930}) with correlation time roughly equal to the largest-eddy turnover time. The forcing is applied with equal power between wavenumbers $kL/2\pi = 2$ and $kL/2\pi = 4$. We perform implicit large-eddy simulations (ILES) without explicit viscosity or resistivity, i.e. the simulations rely on numerical dissipation (this is not true for the dynamo simulation which uses explicit dissipation, see below). Such ILES are found to match the results from direct numerical simulations (DNS) for properly chosen explicit diffusion coefficients (\citealt{Grete:2023}). We consider both fully solenoidal driving and solenoidal + compressible driving, i.e., the driving force creates a mix of curl-free and divergence-free velocity perturbations, $\delta u = (1-f_{\rm s}) \delta u_{\rm comp} +f_{\rm s} \delta u_{\rm sol}$, where $f_{\rm s}$ is the parameter that quantifies the relative fraction of solenoidal and compressive driving ($f_s=1$ corresponds to purely solenoidal driving). The partially compressible simulations are motivated by the fact that compressible MHD fast modes are believed to be important scatterers of CRs by a significant fraction of the CR transport community (e.g, \citealt{yan_lazarian_2004}; \citealt{yan_lazarian_2008}; \citealt{fornieri_2021}). The turbulent energy injection rate is chosen such that all net-flux simulations are subsonic with rms turbulent velocity $ u_{\rm rms}\approx 0.5c_s$, where $c_s$ is the isothermal sound speed. The simulations are evolved for 5 outer-scale eddy turnover times ($\approx$10 box sound crossing times), which ensures steady state in all statistical properties of the turbulence. The initial magnetic field is uniform and along the box diagonal, i.e. $B_{0x} = B_{0y} = B_{0z}$. The simulations are initialised with either $\beta_0 = 100$ or $\beta_0=1000$, where $\beta_0 = 8 \pi p_g/B_0^2$ is the initial ratio of gas pressure to magnetic pressure. The chosen $\beta_0$ values are large enough that the turbulence produced is super-Alfv\'enic with $\delta B/B_0 \approx 4$ in the $\beta_0=100$ simulation and $\delta B/B_0 \approx 10$ in the $\beta_0=1000$ simulation. Consistent with the results of the subsonic, super-Alfv\'enic simulations in \cite{Grete:2021}, the magnetic and kinetic power spectra show scalings close to $k^{-5/3}$ and $k^{-4/3}$, respectively (not shown in this paper). While $\beta_0$ is the input parameter in our simulations, what actually matters for CR transport is the root-mean-square (rms) fluctuation amplitude $\delta B/B_0$ generated by the turbulence. For this reason, we will typically use the value of $\delta B/B_0$ to distinguish between the different mean-field simulations.

We complement these net-flux simulations with an MHD simulation of dynamo at high magnetic Prandtl number (Pm; defined as the ratio of fluid viscosity to magnetic resistivity) taken from \cite{alisa_dynamo}, which we use to gain insight into the physics of particle transport in laminar magnetic folds. This simulation serves as a simplified setup for studying the impact of field reversals and bends on particle transport, although the smooth flow implied by the small Reynolds number ($\approx 20$, due to the high Pm and limited achievable scale separation in the simulation) is not representative of the turbulence in the ISM.  For details about the high-Pm dynamo simulation, we refer the reader to \cite{alisa_dynamo}. Its parameters are provided, along with those of our net-flux simulations, in Table \ref{tab:sims}. 

\begin{figure*}
  \centering
    \includegraphics[width=\textwidth]{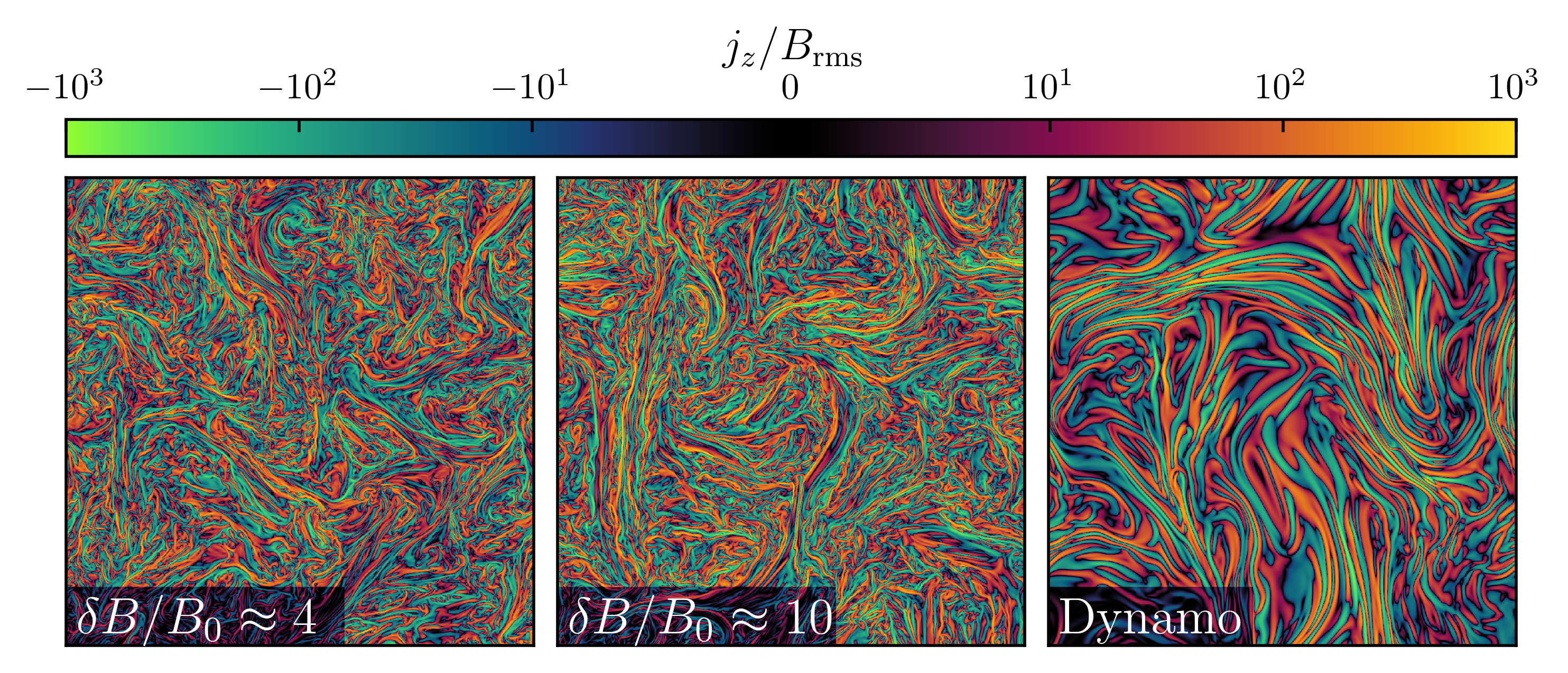}
   \caption{ 2D spatial slices of current density in the simulations used in this work (see Table~\ref{tab:sims}). The turbulence in both mean-field simulations is super-Alfv\'enic ($\delta B/B_0\approx 4$ and $\delta B/B_0 \approx 10$) and produces highly intermittent tangled magnetic fields with small-scale field reversals and current sheets. In the high-Pm dynamo simulation the flow is not very turbulent due to a low Reynolds number (${\rm Re \sim 20}$; \citealt{alisa_dynamo}) and the elongated current sheets are much more laminar and coherent. \label{fig:jz}}
\end{figure*}

We show 2D spatial slices of the current density, $\bb{j} = \grad \btimes \bb{B} $, from the mean-field and dynamo simulations in Figure \ref{fig:jz}. The mean-field turbulence produces highly intermittent tangled magnetic fields with small-scale current sheets. In the high-Pm dynamo simulation the flow is not very turbulent due to a small Reynolds number (${\rm Re} = u_{\rm rms} L / \chi \sim 20$, where $\chi$ is the explicit viscosity used in the simulation; \citealt{alisa_dynamo}) and the current sheets are much more laminar with larger aspect ratios as they are not affected by tearing (\citealt{alisa_dynamo}).

\begin{table}
	\centering
  \begin{adjustbox}{width=0.47\textwidth}
\begin{tabular}{ccccccc}\hline\hline
Sim. Type & $\beta_0$& Resolution & $f_{\rm s}$ & Pm &  $ u_{\rm rms}/c_s$ & $\delta B/B_0 $ \\
\hline\hline
\\
Mean B & $100$ & $1024^3$ & 1 & ---  & 0.5 & $\approx 4$  \\
Mean B& $100$ & $1024^3$ & 0.5 & ---  & 0.5 & $\approx 4$  \\
Mean B& $1000$ & $1024^3$ & 1 & ---  & 0.5 & $\approx 10$  \\
\\
Dynamo& $5 \times 10^{5}$ & $1120^3$& 1 & 500  & 0.18 & $\infty$
\\
\hline
\hline
\end{tabular}

\end{adjustbox}
\caption{Summary of simulations used in this work. $f_{\rm s}$ denotes the fraction of the driving that is solenoidal. The high-Pm dynamo simulation is taken from \citet{alisa_dynamo}.  }\label{tab:sims}  
\end{table}

\subsection{Particle propagation}
We use the Boris-push method (\citealt{boris}) to evolve the Lorentz-force equation,
\begin{equation} \label{eq:lorentz}
     \D{t}{\bb{v}} = \frac{q }{mc} \ \bb{v}\btimes\bb{B}(\bb{r}),
\end{equation}
where $m = \gamma m_0$ is the relativistic mass of a particle with rest mass $m_0$ (e.g., the proton rest mass) and $\gamma$ is the relativistic Lorentz factor. $\bb{B}(\bb{r})$ is the local magnetic field generated by the turbulence at location $\bb{r} = (x,y,z)$. Throughout this work, we use $\bb{v}$ to denote the particle velocities and $\bb{u}$ to represent the velocity of the fluid elements. We do not include electric fields in the Lorentz force as pitch-angle scattering of CRs (e.g., by Alfv\'en waves, fast modes, etc.) is primarily due to magnetic-field fluctuations, while the contribution to scattering from electric fields is negligible for CR speeds much greater than turbulent eddy speeds. In particular, in non-relativistic turbulence with Alfv\'en and sound speeds much smaller than the speed of light, particle acceleration (rate of change of energy) in turbulence is much slower than particle scattering (rate of change of propagation direction) by magnetic-field fluctuations. The acceleration or deceleration of CRs by turbulence is beyond the scope of this work (which in ideal MHD with $\bb{E} = -\bb{u}\btimes\bb{B}/c$ would require time-dependent fields). Our primary focus is the impact of a highly-tangled magnetic field on the spatial and pitch-angle diffusion of particles. 

All particles in our simulations have the same charge $q=1$ and the same speed, $|\bb{v}|\simeq c=1$, i.e., we consider ultra-relativistic CRs. We study the energy dependence of transport by considering different particle masses $m = \gamma m_0$. We use the following notation for the cosine of a particle's pitch angle relative to the magnetic field,
\begin{equation}
    \mu = \frac{\bb{v}\bcdot\bb{B}}{v B},
\end{equation}
and for the particle's magnetic moment,
\begin{equation}
    \mu_M = \frac{m v_\perp^2}{B} = \frac{\gamma m_0 v^2 (1-\mu^2)}{B}. 
\end{equation}
$\mu_M$ is an adiabatic invariant of particle motion assuming fluctuations whose Doppler-shifted wave frequencies are much smaller than the particle gyro-frequency (no gyro-resonant interactions) and whose amplitudes on the gyro-radius scale correspond to energies much smaller than the particle energy (no stochastic heating). 

Instead of using the particle mass $m$ to distinguish between particles of different energies, we will often use the particle gyro-radius (evaluated using the rms magnetic-field strength and $\mu \approx 0$) to describe different-energy particles,
\begin{equation}
    r_{L0} \equiv r_{\rm L} (B_{\rm rms}) = \frac{\gamma m_0 c^2}{qB_{\rm rms}} = \frac{c}{\Omega(B_{\rm rms})}.
\end{equation}
where $\Omega$ is the relativistic gyro-frequency. Higher-energy particles (larger masses $m=\gamma m_0$) correspond to larger gyro-radii. We note that the gyro-radius of the energetically important $\sim$GeV protons is of order $1~{\rm au} \sim 10^{-6}~{\rm pc}$ and so roughly 8 orders of magnitude smaller than typical ISM turbulence injection scales. Our ${\sim}1000^3$ turbulence simulations therefore properly resolve the transport of CR protons that have energies ${\gtrsim} 10^5$~GeV. 

In our MHD turbulence snapshots, the magnetic-field values are specified on a three-dimensional grid. To solve equation~\eqref{eq:lorentz}, we use two methods to interpolate the fields to the particles' locations: we use the triangular-shaped-cloud (TSC) interpolation method (\citealt{tsc_ref}) to evaluate (1) the magnetic field or (2) the vector potential at the particle location. We use method (2) primarily to make sure that the interpolation preserves $\grad\bcdot\bb{B} = 0$, so that artificial changes in particle magnetic moment are not introduced by a non-zero divergence in $\bb{B}$ (the magnetic moment, $\mu_M$, is an adiabatic invariant only if $\grad\bcdot\bb{B}=0$ is satisfied). Nevertheless, we generally find that both methods yield essentially the same results in large-amplitude turbulence.

We consider a wide range of particle masses, including a range of masses that correspond to $r_L(B_{\rm rms}) < \Delta x$, where $\Delta x$ is the grid spacing. The gyro-orbits of these particles are nevertheless resolved in parts of the simulation domain due to the fact that the magnetic-field strength varies by orders of magnitude throughout the box. As a result, particle gyro-radii can be less than the grid spacing in parts of the simulation box and much larger than the grid spacing in other regions. For the numerical integration of equation~\eqref{eq:lorentz} we choose a timestep that depends on particle mass to ensure that $\Delta t \ll \Omega^{-1}$. We typically use $\Delta t = {\rm min} [0.05 \Omega(B_{\rm rms})^{-1},  \Delta x / 4c]$ (so that there are typically 125~steps per gyro-orbit); we have verified that we obtain similar results for smaller timesteps. The particles are allowed to cross the periodic boundaries used in the turbulence simulations.

\begin{figure}
  \centering
        \begin{minipage}[b]{\textwidth}
\includegraphics[width=0.48\textwidth]{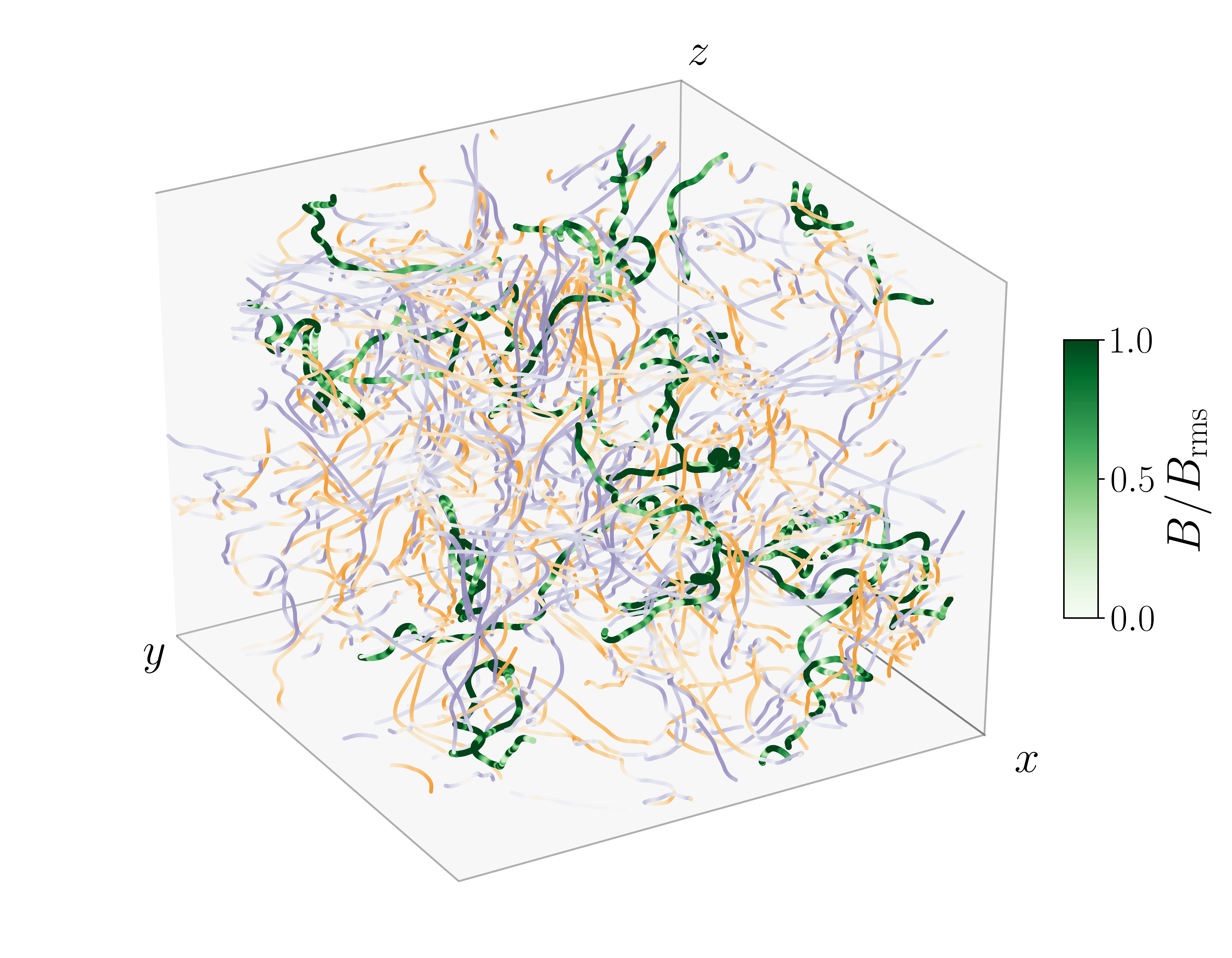}
        \end{minipage} 
        \begin{minipage}[b]{\textwidth}
    \includegraphics[width=0.48\textwidth]{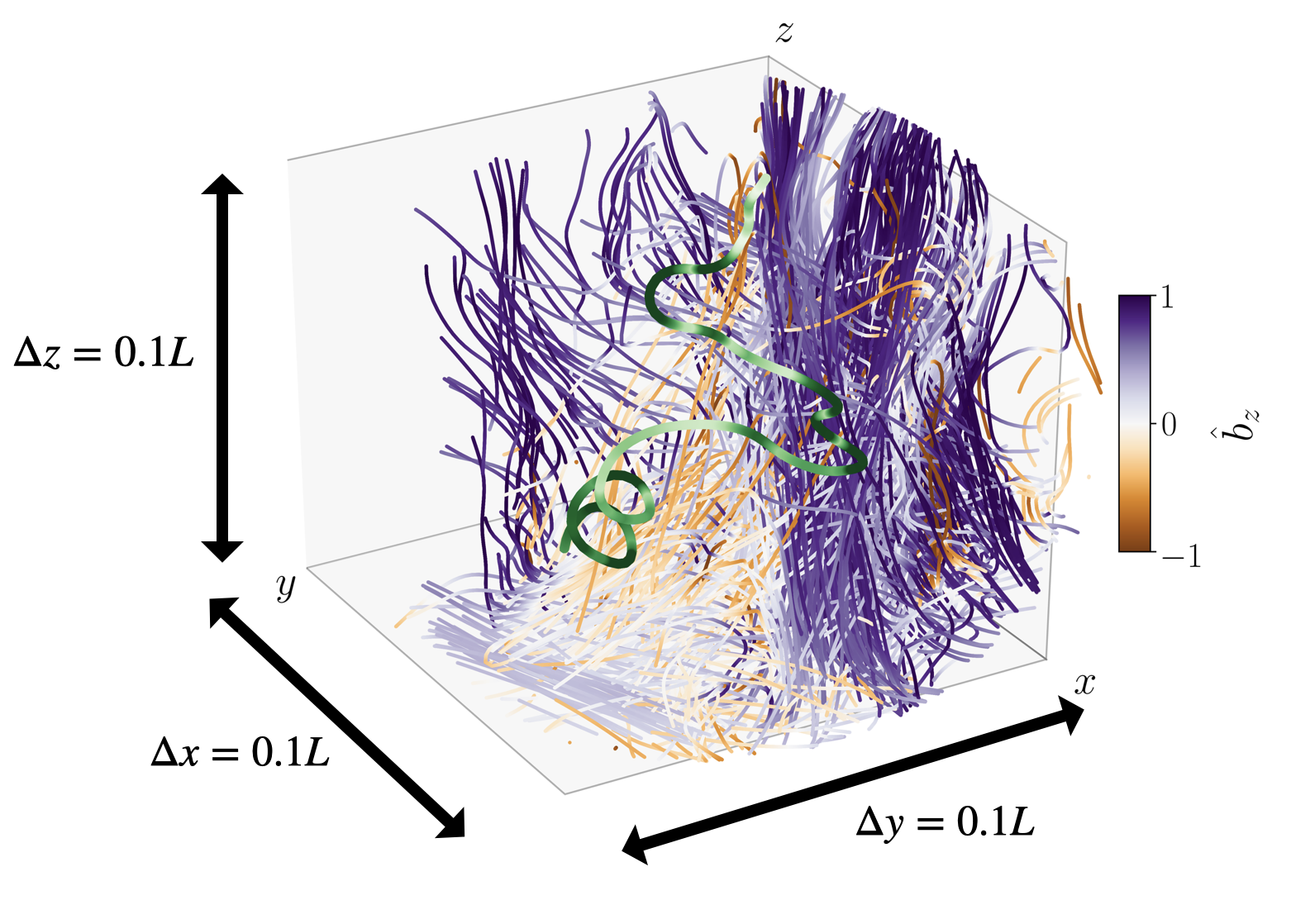}
        \end{minipage} 
   \caption{Top: particle trajectories for $r_{L0}=0.0256 L$ in the $\delta B/B_0 \approx 10$ turbulence simulation (in the full simulation box). The orange-purple lines show the direction of a sample of magnetic field lines (colour-coded using $\hat{b}_z$) while the green lines show the particle trajectories. Unlike the strong guide-field limit, the particles do not follow approximately helical trajectories as they are unable to follow small-scale bends and reversals in the magnetic-field direction. This is despite the fact that their gyro-radii evaluated using $B_{\rm rms}$ are much smaller than the turbulence injection scale. Bottom: single particle trajectory (green) for $r_{L0}=0.0064 L$ in a zoomed-in box. The bottom panel shows that at smaller $r_{L0}$ trajectories are on average more helical although there are again significant local deviations from helical motion in regions of weak $B$ and where field lines reverse, which is important for setting global transport rates. \label{fig:traj_unmag}}
\end{figure}

\subsection{Turbulent leaky box } \label{sec:leaky_box_setup}
In the presence of small-amplitude  fluctuations, i.e.~$\delta B/B_0 \ll 1$, particle pitch angles change at a rate that is small compared to the gyro-frequency. If particles accumulate a change $\Delta \mu$ over a time $\Delta t$, one can estimate the pitch-angle diffusion coefficient using
\begin{equation} \label{eq:Dmumu}
    \kappa_{\mu\mu} = \frac{\langle \Delta \mu^2 \rangle}{\Delta t},
\end{equation}
where the angle brackets denote an average over all particles (e.g., \citealt{beresnyak_2011}; \citealt{xu_2013}; \citealt{mertsch_2020}). This expression is, however, not very suitable when particles are not fully magnetized due to motion in large-amplitude and highly intermittent fields. In such fields, changes in CR pitch angle do not necessarily reflect scattering events in the usual sense. For example, a particle's pitch angle may instantaneously change sign when magnetic-field lines reverse in the direction perpendicular to the local magnetic field on scales  much smaller than the CR gyro-radius (i.e., current sheet thickness ${\ll} r_{L}$), even though the particle trajectory is not necessarily deflected by order unity. The top panel of Figure \ref{fig:traj_unmag} shows such unmagnetized particle trajectories for $r_{L0}=0.0256L$ in the $\beta_0=1000$, $\delta B/B_0\approx 10$ turbulence simulation (see Table~\ref{tab:sims}). The orange--purple lines show the directions of a sample of magnetic-field lines. The field lines are colour coded using $\hat{b}_z$, with dark orange corresponding to $\hat{b}_z=-1$ and dark purple to $\hat{b}_z=1$ (although in the top panel we artificially increase the transparency of the field lines to better show the particle trajectories). The green lines, which show the particle tracks, demonstrate that particle motion is far from helical. The particles do not follow small-scale bends and reversals in the magnetic-field direction despite the fact that their gyro-radii evaluated using $B_{\rm rms}$ are much smaller than the turbulence injection scale. This is partly due to the fact that there are regions in which $B$ is very small (the shade of green in the particle trajectories shows the value of the local magnetic field strength). Particle trajectories become more helical for even smaller $r_L$, as shown in the bottom panel of Figure~\ref{fig:traj_unmag} for $r_{L0}=0.0064$. However, there are again significant local deviations from helical motion in regions of weak magnetic field and where field lines reverse. Equation~\eqref{eq:Dmumu} is therefore not a suitable choice of transport diagnostic.

\begin{figure}
  \centering
      \begin{minipage}[b]{\textwidth}
      \includegraphics[width=0.44\textwidth]{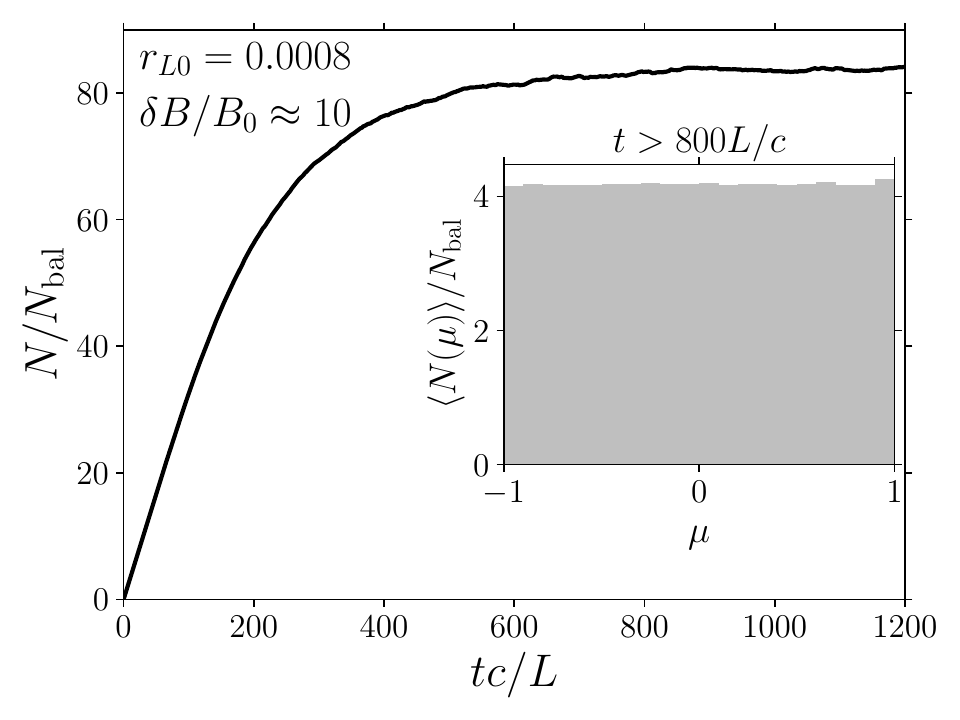}
    \end{minipage} 
      \begin{minipage}[b]{\textwidth}
      \includegraphics[width=0.44\textwidth]{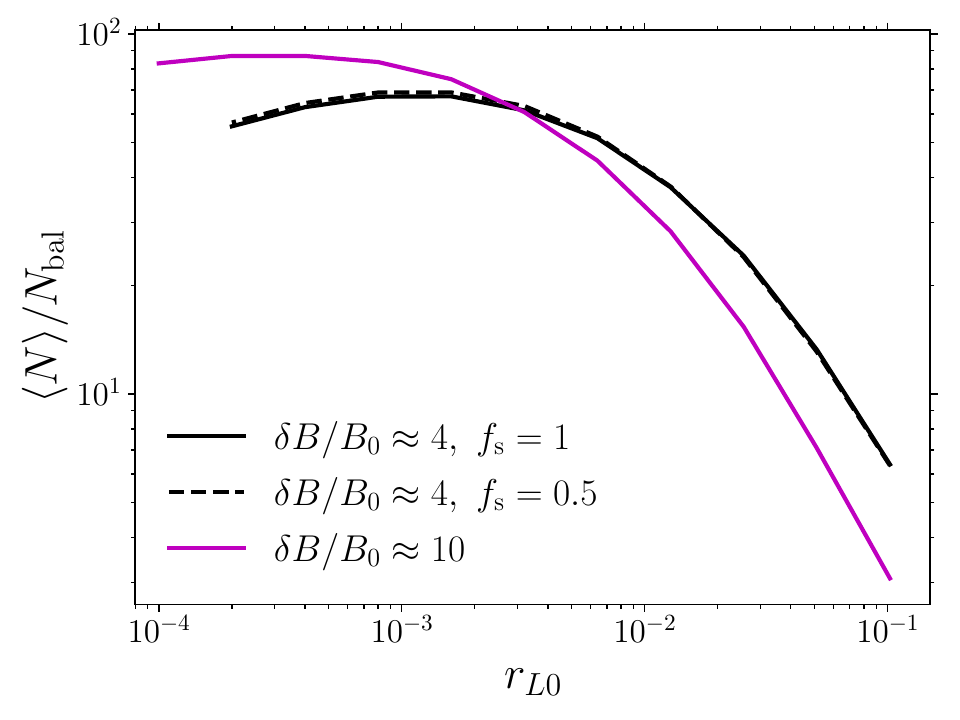}
    \end{minipage} 
  \caption{Turbulent-leaky-box results for the $\delta B/B_0 \approx 4$ and $\delta B/B_0 \approx 10$ turbulence simulations. $N_{\rm bal} \equiv Q H/c$, where $Q$ is the injection rate (the same for all particle energies), is the expected steady-state number of particles for ballistic propagation at the speed of light. The top panel shows that initially the particle number grows, as particles are continuously injected, but have not yet had time to escape (here we use $r_{L0}=0.0008L$ and the $\delta B/B_0 \approx 10$ simulation). Eventually, $N$ reaches a steady state in which injection balances escape. The inset shows that the steady-state, time-averaged pitch-angle distribution of particles present in the box (averaged over $t \in [800,1200]L/c$) is close to isotropic, even though in this simulation we inject particles with a single delta function in pitch-angle, $\mu_0 = 0.99$ (our results are not sensitive to the initial $\mu_0$; see Section \ref{sec:leaky_box_setup} for how particles are initialised). We show the average steady-state particle number as a function of $r_{L0}$ in the bottom panel (averaged over $t \in [800,1200]L/c$, which is after all particle energies reach a steady state). $\langle N \rangle$ increases with decreasing gyro-radius, i.e. lower-energy CRs are better confined, as is the case in the Milky Way. Note that partially compressible driving (dashed line) does not affect the steady-state particle number, indicating that compressive modes in our $\delta B/B_0 \approx 4$ simulations are not important for setting global transport rates.  \label{fig:leaky_box_meanb}}
\end{figure}

Quantifying transport by using a spatial diffusion coefficient $\kappa_{xx}$ of the form  $\kappa_{xx} = \langle \Delta x^2 \rangle / \Delta t$ (\citealt{mertsch_2020}; \citealt{dundovic_2020}; \citealt{hu_2022}) in turn faces the issue that the effective diffusion coefficient has large spatial variations due to intermittency and trapping in magnetic mirrors (e.g., \citealt{xu_2020_trapping}). This complicates the interpretation of $\kappa_{xx}$ measurements, which will depend on whether one considers the median or mean squared displacement, for example (due to the non-Gaussian nature of turbulent fluctuations).  We stress that large variations in local CR transport coefficients are a generic feature of intermittent MHD turbulence and are separate from variations in CR transport on larger scales that may arise from the multi-phase nature of galaxies (\citealt{armillotta_2022}; \citealt{thomas_2022})

 To better probe the energy (or gyro-radius) dependence of CR transport in our simulations, we employ a method that we term the `turbulent leaky box'. In this method, particles are continuously injected throughout the box at fixed time intervals (uniformly in space and constantly in time). We consider particle injection with two types of distributions in initial pitch angle: particles are initialised either with all having the same pitch angle $\mu_0$ or with a uniform distribution in pitch angle. Throughout this work, we show turbulent-leaky-box results for the range of gyro-radii where the choice of initialisation does not matter (because of sufficient scattering, or `optical thickness'). For each particle, we record the initial position vector $(x_0, y_0, z_0)$. We keep track of how far a particle has moved from its injection site, and once it has propagated a distance larger than some prescribed length-scale $H$, i.e., once 
\begin{equation} \label{eq:lb_condition}
    \sqrt{(x-x_0)^2+ (y-y_0)^2 + (z-z_0)^2} > H,
\end{equation}
the particle is said to have escaped and is discarded. This setup generates a steady state number of particles present in the box, which is a direct probe of their volume-averaged effective transport. It incorporates the effect of transport along tangled magnetic-field lines, pitch-angle scattering and trapping between magnetic mirrors (magnetic bottles). $H$ in this setup can be thought of as playing the role of the so-called CR halo size often used in leaky-box models of CR transport in the Galaxy (see e.g., \citealt{Linden2010}; \citealt{blasi12}; \citealt{kq2022} or the review by \citealt{amato_blasi_18}).

If particles random walk due to scattering (not necessarily in pitch-angle, but for example also by following a reversing magnetic-field line) at a rate $\nu_{\rm eff}$, with effective spatial diffusion coefficient $\kappa_{\rm eff} \sim c^2 /\nu_{\rm eff} $, then the steady-state particle number as a function of particle gyro-radius should satisfy,
\begin{equation}
    \langle N(r_{L0}) \rangle \sim Q \tau_{\rm esc} \sim Q \frac{H^2}{\kappa_{\rm eff}}  \propto \nu_{\rm eff}(r_{L0}),
\end{equation}
where $Q$ is the particle injection rate and $\tau_{\rm esc}$ is the characteristic time it takes a particle to escape the turbulent leaky box (see equation~\ref{eq:lb_condition}).
The steady-state number of particles in a turbulent leaky box is therefore a direct probe of the effective spatial diffusion coefficient $\kappa_{\rm eff}(r_{L0}) \propto \nu_{\rm eff}^{-1} \propto (\langle N(r_{L0}) \rangle)^{-1}$.

\section{Turbulent leaky box results} \label{sec:leaky_box_results}
We show our turbulent-leaky-box results for the $\delta B/B_0 \approx 4$ and $\delta B/B_0 \approx 10$ turbulence simulations in Figure~\ref{fig:leaky_box_meanb}. We use $H=4L$ and $H=3L$ respectively, which are chosen to ensure that the steady-state particle distribution function is uniform in pitch angle (using a smaller $H$ runs into the issue of preferentially confining low-energy particles that are trapped in magnetic bottles). Using $H>L$ means that particles can cross the box (which has periodic boundary conditions) and so there is the potential issue that they will sample the same fluctuations more than once. However, due to the highly tangled nature of the magnetic field in the large-amplitude simulations, the probability of returning to the same location after crossing the box is small, and so this issue is not as severe as in simulations with a strong guide field.  

We normalise the particle number $N$ using $N_{\rm bal} \equiv Q H/c$, where $Q$ is the injection rate, which is the same for all particle energies. $N_{\rm bal}$ is therefore the expected steady-state number of particles if particles propagate in a ballistic fashion. For every particle gyro-radius, we inject a total of 76800 particles over the course of the leaky-box simulation. In the top panel, we show how the number of particles $N$ with $r_{L}(B_{\rm rms})=0.0008L$ in the leaky box changes over time for the $\delta B/B_0 \approx 10$ turbulence simulation. Initially, the particle number grows, as particles are continuously injected, but have not yet had time to escape. Eventually, $N$ reaches a steady state in which injection balances escape. The inset shows the steady-state, time-averaged pitch-angle distribution of particles present in the box (averaged over $t \in [800,1200]L/c$).\footnote{Steady states in the leaky box are achieved over timescales that are much larger than $ L/c$ and so it is at first not entirely clear if our use of stationary MHD snapshots remains well motivated. However, for typical ISM conditions the sound and Alfv\'en speeds are orders of magnitude smaller than the speed of light, e.g., $c_s / c \sim 3 \times 10^{-5}$ at temperatures $T\sim 10^4$~K. For our simulations with $ u_{\rm rms}/c_s \sim 0.5$, this means that the outer scale eddy turnover time is ${\approx} 10^5 L/c$ at $T\sim10^4$~K or ${\approx} 10^4 L/c$ at $T\sim10^6$~K. While the eddy turnover time on smaller scales is shorter (e.g., the cascade time in Kolmogorov-like turbulence scales as $t_{\rm casc}(k) \propto k^{-2/3}$), particle transport via advection in turbulence is likely dominated by large-scale motions.} This distribution is close to isotropic, despite the fact that in this particular simulation we inject particles with a single delta function in pitch-angle, $\mu_0 = 0.99$ (our results are insensitive to the choice of initial $\mu_0$). 

The bottom panel of Figure~\ref{fig:leaky_box_meanb} shows the average steady-state particle number as a function of $r_{L0}$ for the $\delta B/B_0 \approx 4$ and $\delta B/B_0 \approx 10$ turbulence simulations ($N$ is averaged over $t \in [800,1200]L/c$, i.e. times for which we find that $N$ is in a steady state for all particle energies). $\langle N \rangle$ increases with decreasing gyro-radius, i.e. lower-energy CRs are better confined, as is the case in the Milky Way. Driving with $f_s=1/2$ (dashed line) does not affect the steady-state particle number  in the simulations with $\delta B/B_0 \approx 4$. This is because compressive modes are not setting global transport rates in our simulations (although we note that compressive motions are generally important for the acceleration of particles in turbulence; e.g. \citealt{brunetti_lazarian_2007}; \citealt{lynn_2013}; \citealt{zhdankin_2021}). In fact, even with $50\%$ compressible driving, the kinetic energy in compressive motions is only of order $1\%$ of the total kinetic energy of the turbulence (see also \citealt{makwana_yan_2020} and \citealt{Gan_2022} for examples of inefficient generation of fast modes by compressive driving), likely due to a combination of dissipation of compressive modes via steepening and compressions generating solenoidal motions through the magnetic tension force.  At the smallest gyro-radii, $\langle N \rangle$ is approximately constant in the $\delta B/B_0 \approx 10$ case, with a more pronounced turnover in the $\delta B/B_0 \approx 4$ case. We discuss the turnover in more detail in Section \ref{sec:net_flux}.

\begin{figure}
  \centering
    \includegraphics[width=0.49\textwidth]{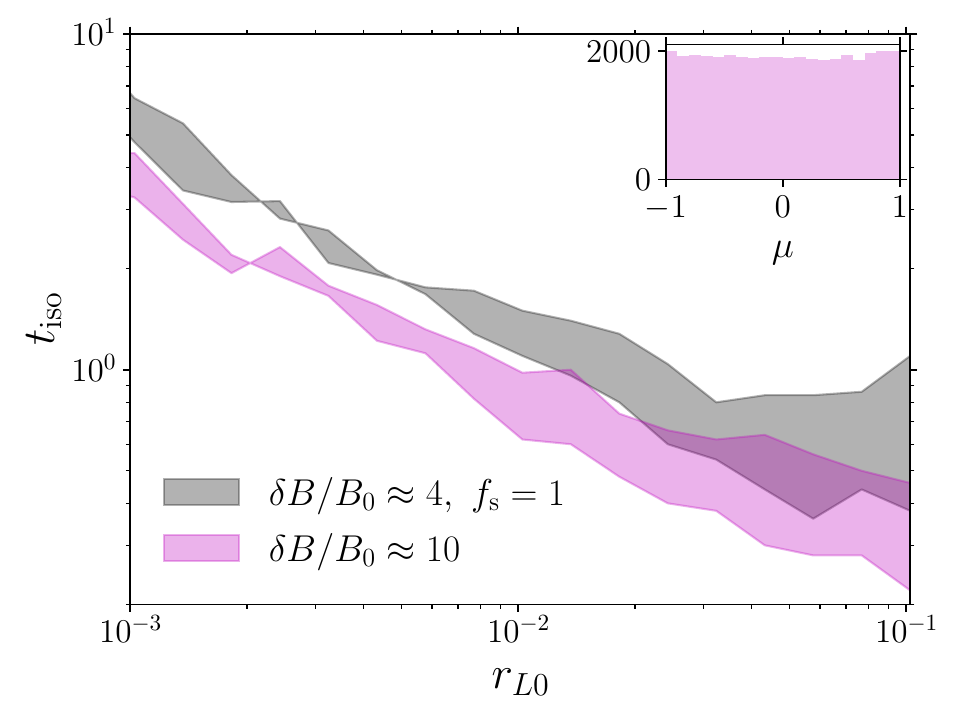}

  \caption{ Pitch-angle isotropisation time, $t_{\rm iso}$, as a function of gyro-radius for the net-flux $\delta B/B_0 \approx 4$ and $\delta B/B_0 \approx 10$ simulations. $t_{\rm iso}$ is defined as the time it takes a distribution of particles initialised with the same pitch angle $\mu_0$ to relax to a uniform distribution in $\mu$. The shaded regions indicate the span of estimated $t_{\rm iso}$ for $\mu_0=0.1$ and $\mu_0=0.9$. We classify the distribution as isotropic once the $\mu$ histogram deviates from being perfectly flat by less than the associated Poisson noise, i.e. $\sum_{i=1}^{N_{\rm bins}} (N_i-N/N_{\rm bins})^2 < N$, where $N_i$ is the number of particles in the $i^{\rm th}$ $\mu$-bin, $N$ is the total number of particles  and $N_{\rm bins}$ is the number of bins. The inset shows the distribution classified as isotropic in the $\delta B/B_0 \approx 10, \mu_0=0.9$ case at the smallest $r_{L0}$. Low-energy particles are isotropised in pitch-angle slower than high-energy particles, which is in stark contrast to Figure \ref{fig:leaky_box_meanb}, which shows that low-energy particles are better confined.  \label{fig:t_iso}}
\end{figure}

\begin{figure}
  \centering
      \begin{minipage}[b]{\textwidth}
      \includegraphics[width=0.45\textwidth]{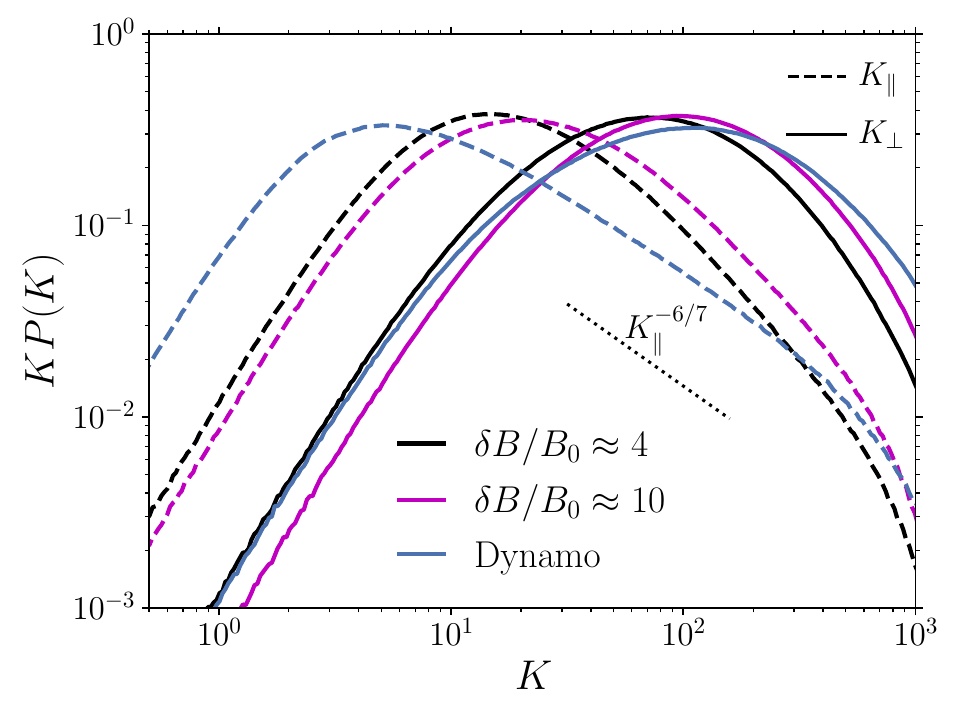}
    \end{minipage} 
      \begin{minipage}[b]{\textwidth}
      \includegraphics[width=0.45\textwidth]{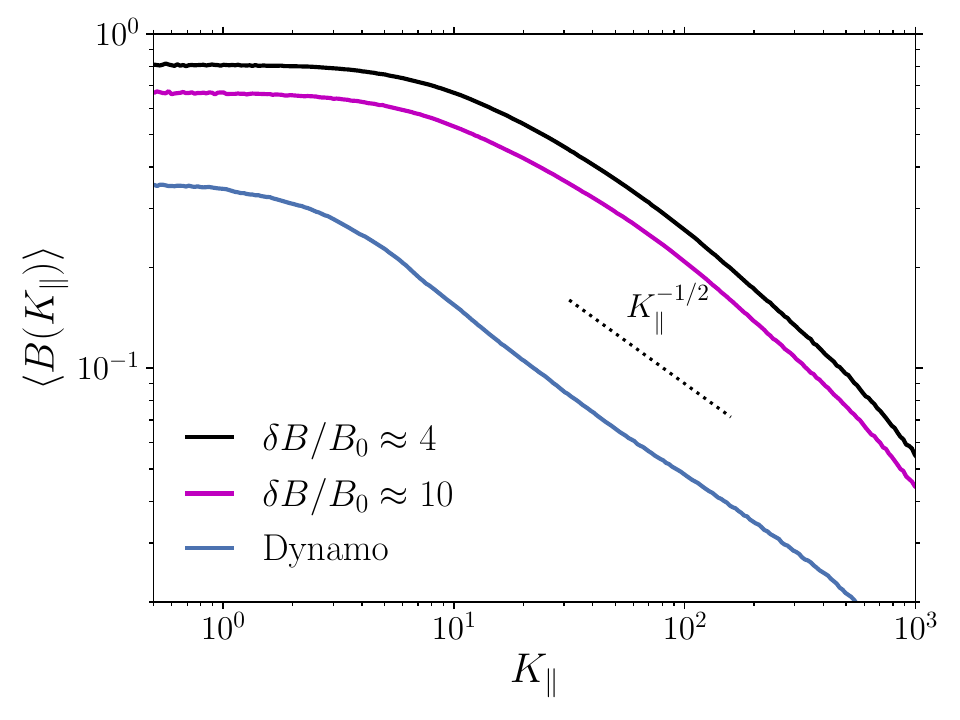}
    \end{minipage} 
  \caption{Probability density functions (PDFs) of the field-line curvature $K_\parallel = |\hat{\bm b} {\bm \cdot \nabla} \hat{\bm b}|$ (dashed lines; computed numerically by also removing its component along $\hat{\bm b}$, i.e. $K_\parallel = |\hat{\bm b} {\bm \times} (\hat{\bm b} {\bm \cdot \nabla} \hat{\bm b})|$) and inverse perpendicular reversal scale $K_\perp = |\hat{\bm b} \btimes (\hat{\bm b} \btimes \bm{\nabla} \ln B )|$ (solid lines). The PDFs are multiplied by $K$ (either $K_\parallel$ or $K_\perp$, depending on the PDF) so that the peaks show the characteristic curvature and reversal scales. The black and magenta lines show the results for the $\delta B/B_0 \approx 4$ and $\delta B/B_0 \approx 10$ mean-field simulations, respectively, while the blue line shows the result for the ${\rm Pm}=500$ dynamo. For all simulations, the $K_\perp$ PDF peaks on smaller scales than the $K_\parallel$ PDF, which means that magnetic fields are organized into structures that resemble folds with small perpendicular coherence lengths and larger parallel coherence lengths. The scale separation between the peaks is particularly large for the high-Pm dynamo, which has a small Reynolds number, ${\rm Re} \approx 20$, and so it is not very turbulent. In this case the folds are much more laminar and elongated along the local magnetic field. In the bottom panel, we show that regions of large curvature have smaller than average magnetic field strengths, consistent with the scaling $B\propto K_\parallel^{-1/2}$ often found in MHD dynamo simulations (e.g. \citealt{schekochihin_2004}), which corresponds to magnetic fields being organized into structures that keep the magnetic tension force $\bb{B}\bcdot\grad \bb{B} \approx {\rm const}$.
\label{fig:pdfs}}
\end{figure}

To understand the results from our turbulent-leaky-box experiments, it is important to realise that the transport of CRs in our simulations is set by physics that is very different from pitch-angle diffusion in the strong guide field limit. This becomes apparent by looking at Figure \ref{fig:t_iso}, which shows the pitch-angle isotropisation time, $t_{\rm iso}$, as a function of gyro-radius for the $\delta B/B_0 \approx 4$ and $\delta B/B_0 \approx 10$ simulations. We define $t_{\rm iso}$ as the time it takes a distribution of particles initialised with the same pitch-angle $\mu_0$ to relax to an approximately uniform distribution in $\mu$. The shaded regions indicate the span of estimated $t_{\rm iso}$ for $\mu_0=0.1$ and $\mu_0=0.9$. We classify the sampled $\mu$ histogram as isotropic once it deviates from being perfectly flat by less than the associated Poisson noise, i.e. $\sum_{i=1}^{N_{\rm bins}} (N_i-N/N_{\rm bins})^2 < N$, where $N_i$ is the number of particles in the $i^{\rm th}$ $\mu$-bin, $N$ is the total number of particles  and $N_{\rm bins}$ is the number of bins. The inset shows the distribution classified as isotropic in the $\delta B/B_0 \approx 10, \mu_0=0.9$ case at the smallest $r_{L0}$.

In the traditional picture of particle transport with a strong guide field, the results from Figure \ref{fig:t_iso}, which show that low-energy particles are isotropised in pitch-angle slower than high-energy particles, would be in stark contrast to Figure \ref{fig:leaky_box_meanb}, which demonstrates that lower-energy particles are better confined. In particular, within the standard paradigm of pitch-angle diffusion in the presence of a strong guide field, one expects $\langle N \rangle \propto t_{\rm iso}^{-1}$, since the spatial diffusion rate is inversely proportional to the pitch-angle diffusion rate. This is very different from what we find in the case of a weak guide field. This suggests that particle transport is not set by the rate of pitch-angle diffusion in the usual sense and that we need a different transport description.

\section{Model For particle transport in reversing magnetic fields} \label{sec:model}

Before we attempt to find an appropriate new model, we first identify key statistical quantities that probe the intermittency of MHD turbulence and which turn out to affect particle transport in our simulations significantly.

\subsection{Turbulence statistics in the weak-guide-field limit}
In the standard quasi-linear treatment of CR transport in turbulence with a strong guide field, the pitch-angle diffusion coefficient is set by the magnetic-field power spectrum. The power spectrum, however, does not contain information about intermittency and sharp features in the fields, which are believed to be ubiquitous in MHD turbulence, especially when $\delta B/B_0 \gg 1$. For turbulence simulations that have $\delta B /B_0 \gg 1$, which are the focus of this paper, it is therefore more instructive to work with statistical quantities usually considered in the context of the MHD dynamo (e.g., \citealt{schekochihin_2004}; \citealt{rincon_dynamo}; \citealt{alisa_dynamo}). In particular, it is useful to consider length-scales associated with order-unity changes in the magnetic field. To study variations along the field, we consider the field-line curvature,
\begin{equation} \label{eq:kpar}
    \bb{K}_\parallel \equiv  \eb\bcdot\grad\eb,
\end{equation}
where $\eb\equiv\bb{B}/B$ is the unit vector directed along the local magnetic-field direction. To probe variations across the field, the dynamo literature often considers (e.g., \citealt{schekochihin_2004}; \citealt{alisa_dynamo})
\begin{equation} 
    \bb{K}_{\bs{j\times B}} \equiv \frac{\bb{j}\btimes\bb{B}}{B^2} = \frac{(\grad\btimes\bb{B})\btimes\bb{B}}{B^2} .
\end{equation}
In this work, we will use a related but slightly different probe of the perpendicular reversal scale,
\begin{equation} \label{eq:kperp}
    \bb{K}_\perp \equiv \bb{K}_{\bs{j\times B}} - \bb{K}_\parallel = \eb\btimes(\eb\btimes\grad\ln B).
\end{equation}
We refer to $K_\perp^{-1}$ as the field-line perpendicular reversal scale. Both $K_\parallel$ and $K_\perp$ have significant implications for particle transport. When $K_\parallel \sim r_{L}^{-1}$, i.e. the field line bends on a scale comparable to the \textit{local} gyro-radius, one expects the particle to leave the field line and/or undergo a strong scattering event with order-unity change in pitch angle. On the other hand, if $K_\perp \sim r_{L}^{-1}$ the particle is unable to follow the field line as its $\nabla B$-drift perpendicular to the magnetic field becomes comparable to its speed along the field line.    

\begin{figure}
  \centering
  \centering
      \begin{minipage}[b]{\textwidth}
      \includegraphics[width=0.48\textwidth]{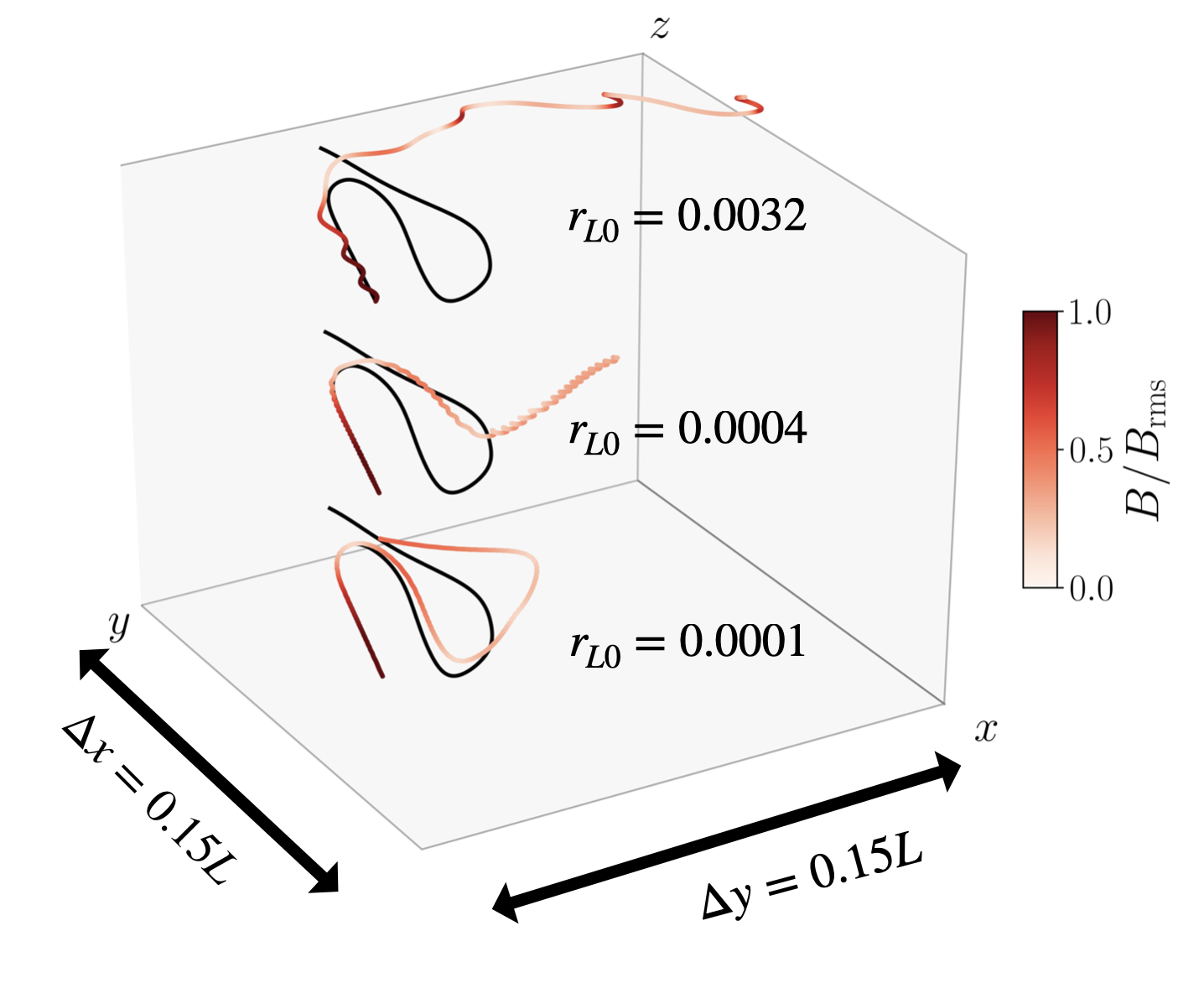}
    \end{minipage} 
      \begin{minipage}[b]{\textwidth}
      \includegraphics[width=0.48\textwidth]{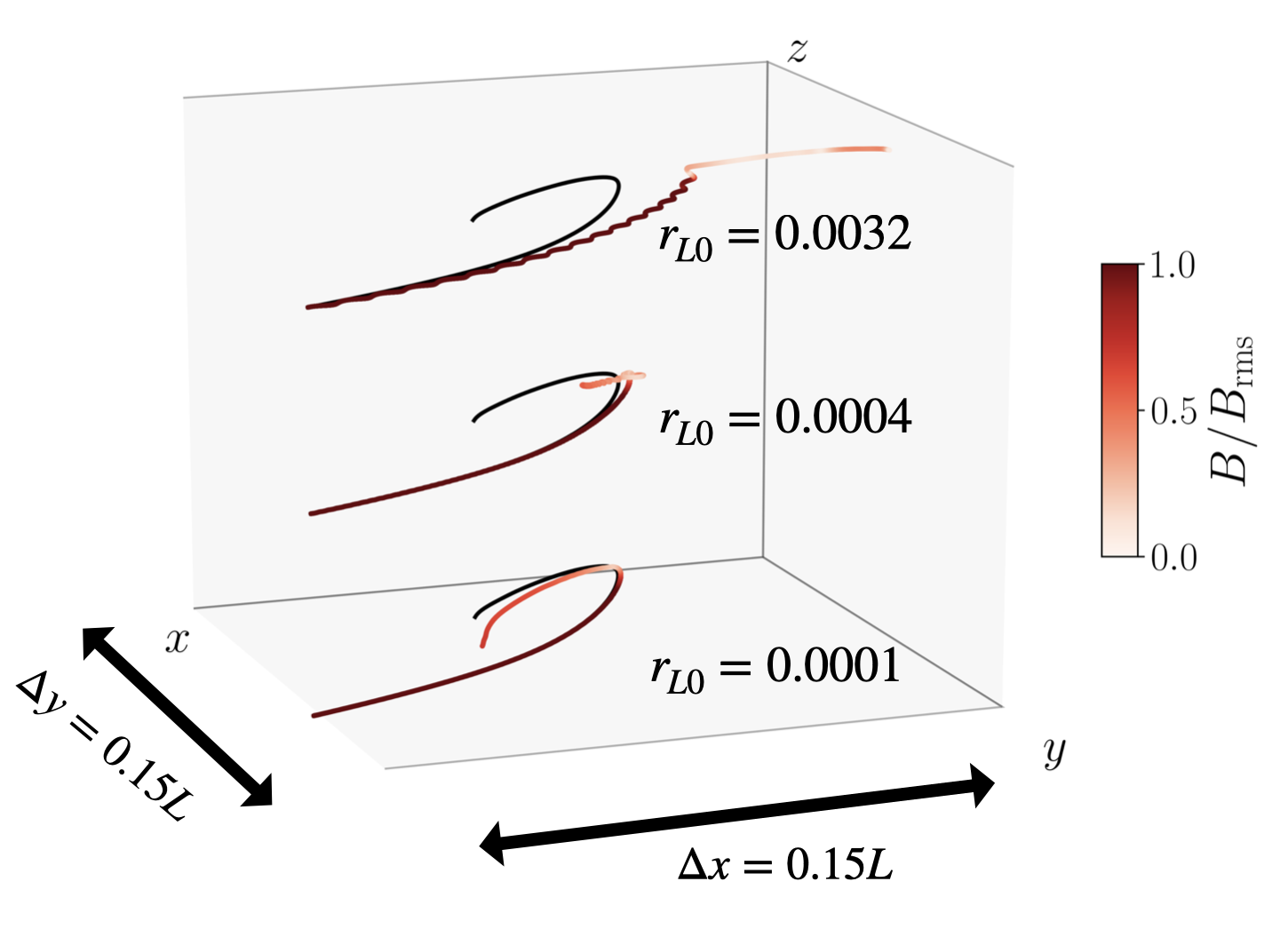}
    \end{minipage} 
  \caption{Example particle trajectories which illustrate propagation in a magnetic fold, taken from the ${\rm Pm=500}$ dynamo simulation. In both panels, the black lines all show the same magnetic field line, but we have displaced the field lines and particle trajectories for visualisation purposes (here we use $r_{L0} = 0.0001L, \ 0.0004L, \ 0.0032L$ and the displaced  particle trajectories start from the same location). The particle trajectories are shown in red and the shade of red indicates the strength of the local magnetic field. As particles propagate along a fold and approach the location where the field bends (high curvature), the gyro-radius increases due to the decrease in magnetic field strength (see Figure \ref{fig:pdfs}). Due to this increase in local gyro-radius and decrease in local gyro-frequency, only the $r_{L0}=0.0001L$ particle is able to follow the folded field line as it bends reasonably well. The $r_{L0}=0.0032L$ particle drifts out of the fold before the field-line bends due to large $\nabla B$-drift. In the bottom panel the intermediate-mass particle is able to propagate further into the fold but eventually gets scattered by the high curvature of the bending field line, where the local gyro-frequency is of order $K_\parallel c$.    \label{fig:fold_trajectories}}
\end{figure}

We show probability density functions (PDFs) of the field-line curvature ($K_\parallel$, dashed lines) and inverse perpendicular reversal scale ($K_\perp$, solid lines) evaluated using our simulations in Figure \ref{fig:pdfs}. The PDFs are multiplied by $K$ (either $K_\parallel$ or $K_\perp$, depending on the PDF) so that the peak corresponds to the scale that dominates the total probability. The black and magenta lines show the results for the $\delta B/B_0 \approx 4$ and $\delta B/B_0 \approx 10$ mean-field simulations, respectively, while the blue line shows the result for the ${\rm Pm}=500$ dynamo. In all cases, the $K_\perp$ PDF peaks on smaller scales than the $K_\parallel$ PDF, an indication that magnetic fields are organized into structures that resemble folds with small perpendicular coherence lengths and larger parallel coherence lengths. Such structures are commonly seen in the context of the MHD dynamo (\citealt{schekochihin_2004}; \citealt{rincon_dynamo}; \citealt{alisa_dynamo}). The scale separation between the peaks of the $K_\parallel$ and $K_\perp$ distributions is smaller in the guide-field turbulence simulations than in the high-Pm dynamo (blue line); this is because the high-Pm dynamo has a small Reynolds number, ${\rm Re} \approx 20$, and so it is not very turbulent. In this case the folds are more laminar and less affected by tearing (\citealt{alisa_dynamo}), which results in them being more elongated along the local magnetic field. We also note that, at large $K_\parallel$, $P(K_\parallel)$ in the ${\rm Pm}=500$ dynamo simulation shows a scaling that is close to the $K_\parallel^{-13/7}$ scaling predicted for the MHD dynamo (\citealt{schekochihin_2001}; \citealt{schekochihin_2004}). For the simulations with a net magnetic field, in which the magnetic field is bent by turbulence rather than a relatively smooth low-Re flow, $P(K_\parallel)$  forms a steeper power law. In the bottom panel we show the average magnitude of the magnetic field as a function of the field-line curvature. For all simulations, regions of large curvature have smaller than average magnetic field strengths. The scaling at large $K_\parallel$ is consistent with the  $B\propto K_\parallel^{-1/2}$ scaling often found in  MHD dynamo simulations (e.g. \citealt{schekochihin_2004}), which corresponds to magnetic fields being organized into structures that keep the magnetic tension force $\bb{B}\bcdot\grad\bb{B} \approx {\rm const}$.

\subsection{Insights from the high-Pm dynamo: transport in laminar magnetic folds}
Because magnetic folds created by a small-Re, high-Pm dynamo are very smooth, the ${\rm Pm=500}$ simulation is quite useful for gaining intuition regarding particle transport in reversing magnetic fields. In particular, because the ${\rm Pm=500}$ dynamo is not very turbulent due to its small Reynolds number, it allows us to study particle transport in folds without the complexity that would be added by the additional presence of turbulent fluctuations. We stress that the high-Pm dynamo simulation used here is not meant to mimic fully turbulent systems and its primary application is to study how the statistics of field reversals and bends affect particle transport.   

In Figure~\ref{fig:fold_trajectories}, we show example trajectories of particles with 3 different rms gyro-radii ($r_{L0} = 0.0001L, \ 0.0004L, \ 0.0032L$) measured near two different magnetic folds taken from the ${\rm Pm=500}$ dynamo simulation. In both panels, the three black lines trace the same magnetic-field line, but we have displaced the field lines and particle trajectories (which all start from the same location) for visualisation purposes. The particle trajectories are shown in red and the shade of red indicates the strength of the local magnetic field. As particles propagate along a fold and approach the location where the field bends (high curvature), the gyro-radius increases due to the decrease in magnetic-field strength (see Figure~\ref{fig:pdfs}). Due to this increase in local gyro-radius and decrease in the local gyro-frequency, only the  $r_{L0}=0.0001$ particle is able to follow the folded field line as it bends. The $r_{L0}=0.0032$ particle drifts out of the fold before the field-line bends due to the large $\nabla B$-drift. In the bottom panel, the intermediate-mass particle is able to propagate further into the fold but eventually gets scattered by the high curvature of the bending field line, where the local gyro-frequency is of order $K_\parallel c$.    

\begin{figure}
  \centering
    \includegraphics[width=0.49\textwidth]{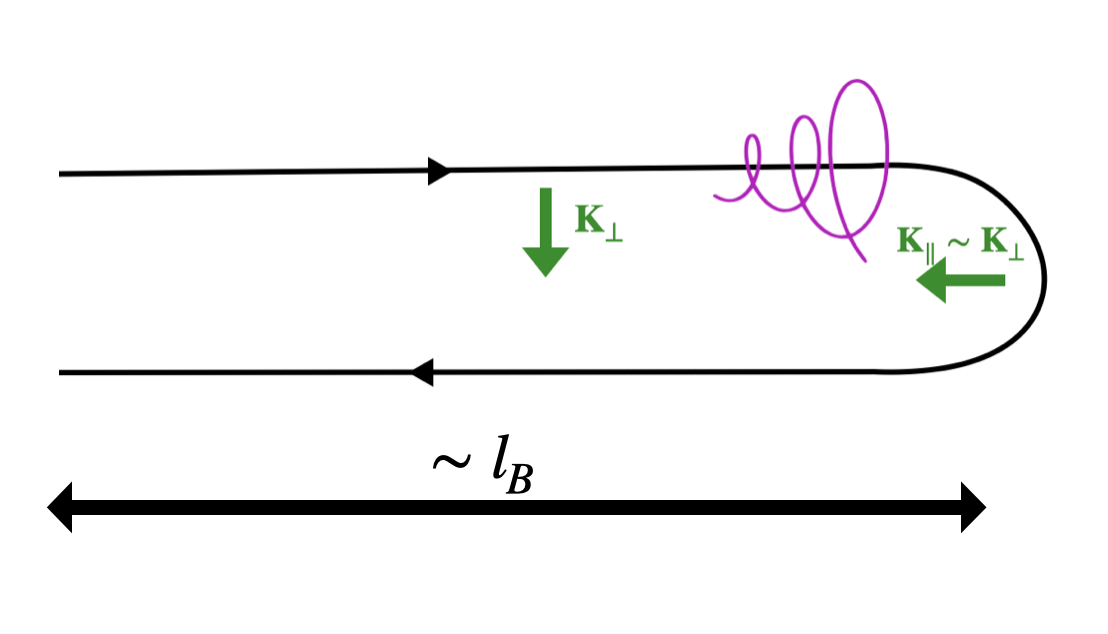}

  \caption{Schematic of particle transport in a magnetic fold that has a parallel coherence length $l_B$ ($\sim$ the scale where the curvature PDF in Figure \ref{fig:pdfs} peaks). $l_B$ is much larger than the perpendicular coherence length $K_\perp^{-1}$, i.e. most of the fold has $K_\parallel \ll K_\perp$. Where the fold reverses, we assume that it has a curvature $K_\parallel$ that is of order $K_\perp$. A particle with gyro-radius $r_{L0} \gg K_\perp^{-1}$ is not magnetized and is not able to follow the field line as it reverses. Instead, it feels a smaller effective magnetic force because it samples many field reversals. As a result, the impact of the magnetic field on the particle trajectory is suppressed. For $r_{L0} < K_\perp^{-1}$, a particle is able to gyrate around the initially $\sim$straight magnetic field line, but as it approaches the region of high curvature the $B$-field drops, the particle's gyro-frequency decreases and its gyro-radius increases (this is illustrated by the schematic magenta trajectory). It can thus fully trace the reversing magnetic field line only if $\Omega(B \sim K_\parallel^{-1/2}) \gg K_\parallel c$. If $\Omega(B \sim K_\parallel^{-1/2}) \sim K_\parallel c$, the particle will experience order unity field-line bending during one gyro-orbit and will be strongly scattered in pitch angle (by order $\sim$ unity). If, however, $\Omega(B \sim K_\parallel^{-1/2}) \ll K_\parallel c$, i.e. the gyro-frequency is too small to tie the particle to the bending field line, the particle will drift out of the fold (see Figure \ref{fig:fold_trajectories}). In a folded magnetic structure a particle that is not scattered in pitch angle and perfectly follows the field line is nonetheless effectively being `scattered' due to the field reversal. \label{fig:fold}}
\end{figure}

\subsection{New model of particle transport in reversing magnetic fields} \label{sec:model_heuristic}
The trajectories from Figure~\ref{fig:fold_trajectories} and the PDFs from Figure~\ref{fig:pdfs} allow us to derive a heuristic model for particle transport in reversing magnetic fields. We show a schematic of what the transport looks like in Figure~\ref{fig:fold}. We consider the following fold geometry: we assume that the fold has a parallel coherence length $l_B$ ($\sim$ the scale where the curvature PDF in Figure~\ref{fig:pdfs} has its peak) that is much larger than the perpendicular coherence length $K_\perp^{-1}$, i.e., most of the fold has $K_\parallel \ll K_\perp$. Where the fold reverses, it has a curvature $K_\parallel$ that is of order $K_\perp$, $K_\parallel \sim K_\perp$. 

Let us now consider particle motion in such a fold. A particle with gyro-radius $r_{L0} \gg K_\perp^{-1}$ is not magnetized and is not able to follow the field line as it reverses. Instead, it feels a smaller effective magnetic force because it samples many field reversals. As a result, the impact of the magnetic field on the particle trajectory is suppressed. If a particle has $r_{L0} < K_\perp^{-1}$, it is able to gyrate around the magnetic-field line where $K_\parallel \ll K_\perp$, but as it approaches the region of high curvature the $B$-field drops, the particle's gyro-frequency decreases and its gyro-radius increases. This is illustrated by the schematic magenta trajectory in Figure~\ref{fig:fold}. The particle is thus able to trace the reversing magnetic field line fully only if $\Omega(B \sim K_\parallel^{-1/2}) \gg K_\parallel c$. If $\Omega(B \sim K_\parallel^{-1/2}) \sim K_\parallel c$, the particle will experience order-unity field-line bending during one gyro-orbit and will be strongly scattered in pitch angle (by a factor of order unity). If, however, $\Omega(B\sim K_\parallel^{-1/2}) \ll K_\parallel c$, i.e. if the particle becomes unmagnetized where the field bends, then the particle will drift out of the fold (see Figure \ref{fig:fold_trajectories}). If $\Omega(B\sim K_\parallel^{-1/2}) \ll K_\parallel c$ at the location of the bend, then it seems plausible that $\Omega(B\sim K_\parallel^{-1/2}) \sim K_\parallel c$ will be satisfied earlier along the particle's trajectory in the fold. This would at first glance suggest significant scattering. However, strong scattering due to “resonant curvature" also requires that the “resonance" is satisfied over length scales ${\sim}K_\parallel^{-1}$, i.e., that a bend corresponds to a field line maintaining curvature $K_\parallel$ over a length scale $K_\parallel^{-1}$. Thus, while the local curvature has a range of values along a fold, most of them do not correspond to actual bends and strong scattering events.  

There is a very important difference between particle transport in such a folded structure and transport in the presence of a strong guide field. In the presence of a strong guide field, a particle that perfectly follows a field line without pitch-angle diffusion does not undergo a random walk. However, in a folded structure a particle that is not scattered in pitch angle and perfectly follows a folded magnetic field is effectively being `scattered' due to the random walk of the magnetic-field lines. Because folds come in a variety of sizes (see the PDF in Figure~\ref{fig:pdfs}), the scattering frequency associated with the ability to follow folded magnetic-field lines depends on the particle energy (or rms gyro-radius). 

With this picture in mind, we can write down a heuristic equation for the effective scattering frequency of a particle  propagating in folded magnetic fields,
\begin{equation} \label{eq:nu}
    \nu (r_{L0}) \sim \frac{c}{l_B} \int^{K_{\rm max}} P(K_\perp) \,\rmd K_\perp,
\end{equation}
i.e. particles are scattered at a rate of order $c/l_B$ (the rate at which they traverse the parallel length of a fold) multiplied by the probability that their local gyro-frequency is sufficiently large and gyro-radius sufficiently small to allow a full traversal of the fold. This condition is encapsulated in the upper integration limit $K_{\rm max}$. The characteristic parallel length of the fold $l_B$ is defined as
\begin{equation} \label{eq:lB}
    l_B = \int K_\parallel^{-1} P(K_\parallel) \,\rmd K_\parallel.
\end{equation}
$K_{\rm max}$ in equation~\eqref{eq:nu} can be estimated using the assumption that the fold bends on a scale comparable to the perpendicular reversal scale, i.e. $K_\parallel \sim K_\perp$ (as shown schematically in Figure \ref{fig:fold}, i.e. the maximum curvature along a fold is comparable to the separation between two oppositely oriented field lines that make up the fold), and that in regions of high $K_\parallel$ there is a smaller-than-average $B$ and so smaller-than-average gyro-frequency and larger-than-average particle gyro-radius (Figure~\ref{fig:fold}). $K_{\rm max}$ is therefore given by the condition that
\begin{equation} \label{eq:Kmax}
    K_{\rm max} c / \Omega(B_{\rm rms})  = K_{\rm max} r_{L0} = f_{r} \, \frac{\langle B(K_\parallel = K_{\rm max}) \rangle}{B_{\rm rms}},
\end{equation}
where $f_r$ is a number of order unity. The exact value of $f_r$ is somewhat unclear without a more detailed microphysical theory and likely depends upon the PDF of $B(K_\parallel)$. On small scales, $\langle B(K_\parallel) \rangle$ is roughly proportional to $K_\parallel^{-1/2}$  (Figure~\ref{fig:pdfs}) and so asymptotically we expect that
\begin{equation} \label{eq:Kmax_asympt}
    K_{\rm max} \propto r_{L0}^{-2/3},
\end{equation}
which can be derived using equation~\eqref{eq:Kmax}.

At first glance, it may seem quite surprising that the scattering rate in equation~\eqref{eq:nu} depends on the PDF of $K_\perp$, since typically perpendicular variations are not important for scattering CRs (e.g., in quasi-linear theory). This is also true for particles propagating in magnetic folds, in the sense that field reversals in the perpendicular direction do not directly scatter CRs. Nevertheless, the perpendicular reversal scale plays an important role as it sets the characteristic length-scale of the bend `at the end' of the fold. The fold's $K_\perp$ therefore sets whether CRs will be scattered after they traverse a distance $l_B$ (coherence length of fold): either due to $K_\parallel c \sim \Omega (B)$, in which case particles undergo order unity pitch-angle scattering (e.g., the $r_{L0}=0.0004$ trajectory in the bottom panel of Figure~\ref{fig:fold_trajectories}); or because $K_\parallel c \ll \Omega(B)$, in which case the particles get `scattered' by following the field as it bends while preserving their pitch angle (e.g. the $r_{L0}=0.0001$ trajectory in the bottom panel of Figure~\ref{fig:fold_trajectories}).

\subsection{Relative importance of field-line random walk and scattering by resonant curvature in setting CR transport }

The upper bound of the integral in equation~\eqref{eq:nu}, i.e. where $K_\parallel c \sim \Omega(B)$, already incorporates order-unity pitch-angle scattering by `resonant curvature' into the overall scattering frequency. Equation~\eqref{eq:nu} therefore includes scattering both due to particle motion along a reversing magnetic field (most of the integration range) and due to pitch-angle scattering by regions of resonant curvature (upper bound of the integral). To quantify which contribution dominates, we first define $K_{\rm peak}$ as the $K_\perp$ for which $K_\perp P(K_\perp)$ in Figure~\ref{fig:pdfs} is maximal.    For large $r_{L0}$ with $K_{\rm max} \ll K_{\rm peak}$ (equation~\ref{eq:Kmax}), the overall scattering frequency is dominated by the resonant curvature contribution according to equation~\eqref{eq:nu} and Figure~\ref{fig:pdfs}. For very small $r_{L0}$ with $K_{\rm max} \gg K_{\rm peak}$, the contribution from resonant curvature is small as the integral in equation~\eqref{eq:nu} is dominated by $K_\perp \sim K_{\rm peak}$. The `scattering' rate in equation~\eqref{eq:nu} is then primarily due to the field-line random walk. Importantly, despite the less efficient pitch-angle scattering by resonant curvature at these small gyro-radii, the overall `spatial scattering' rate is not reduced as the particles' ability to follow reversing field lines is improved. 

Whether field-line random walk or scattering by resonant curvature sets the transport therefore depends on the relative location of $K_{\rm max}$ (which depends on particle energy) and $K_{\rm peak}$.  This in turn depends on the shape of the PDFs in Figure~\ref{fig:pdfs}. The PDFs may look different in higher-resolution simulations and have a larger $K_{\rm peak}$ (see Figure~\ref{fig:pdf_resolution}), thus changing the particle energy at which the transition between the two types of transport occurs.

\subsection{Comparison of fold propagation model to leaky-box results} \label{sec:model_comparison}
We now compare our model for the effective scattering rate (equations~\ref{eq:nu}--\ref{eq:Kmax}) to the leaky box results from Figure~\ref{fig:leaky_box_meanb}. While the heuristic model was inspired by the smooth folds in the high-Pm dynamo simulation, it may also describe the mean-field $\delta B/B_0 \approx 4$ and $\delta B/B_0 \approx 10$ simulations reasonably well, as they are also characterised by perpendicular coherence lengths that are significantly smaller than the parallel coherence lengths (Figure~\ref{fig:pdfs}).  

\begin{figure}
  \centering
    \includegraphics[width=0.49\textwidth]{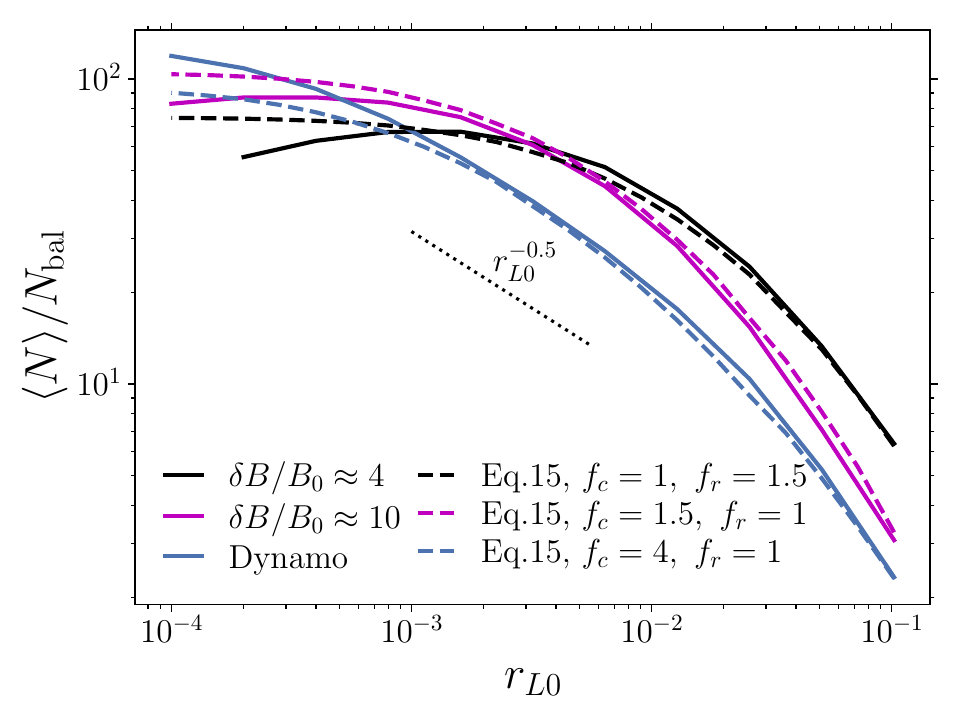}

  \caption{ Fold-propagation-model predictions (eq.~\ref{eq:model_N}, dashed lines) compared to our leaky box results (solid lines), for both the mean-field simulations and the dynamo simulation. Overall there is remarkably good agreement for constants $f_c$ (eq.~\ref{eq:model_N}) and $f_r$ (eq.~\ref{eq:Kmax}) which are close to 1 (for the dynamo $f_c=4$ is somewhat large; we speculate why that is in Section~\ref{sec:model_comparison}). This suggests that the effective transport rate is indeed set by the particles' ability to follow small-scale field line reversals and the scattering by regions of resonant curvature. \label{fig:leaky_box_model}}
\end{figure}

We compare the fold-propagation-model predictions to our leaky-box results in Figure~\ref{fig:leaky_box_model}. We also include leaky-box results for the ${\rm Pm=500}$ dynamo simulation. We estimate the predicted steady-state particle number $N_{\rm m}$ using
\begin{equation} \label{eq:model_N}
     N_{\rm m}(r_{L0}) = f_{c}  \frac{QH^2 \nu(r_{L0})}{\langle { v_\parallel} \rangle ^2} = f_c \frac{QH^2}{\langle { v_\parallel} \rangle  l_{\rm B}}  \int^{K_{\rm max}} P(K_\perp) \rmd K_\perp,
\end{equation}
where $Q$ is the particle injection rate, $\langle { v_\parallel} \rangle=c\langle |\mu|\rangle =0.5 c$ is the expected average speed along a field line, and $l_B$ and $K_{\rm max}$ are calculated using equations~\eqref{eq:lB} and \eqref{eq:Kmax}. $f_c$ is a constant factor that sets the overall normalisation and should preferably be of order unity. Note that $QH^2 \nu(r_{L0} )/ c^2$  in equation~\eqref{eq:model_N} is simply the particle number under the assumption of diffusive propagation with scattering frequency $\nu(r_{L0})$, i.e. $N \sim QH^2/\kappa \sim Q\tau_{\rm esc}$.

Overall there is remarkably good agreement between the model predictions (dashed lines) and the leaky box results (solid lines), for constants $f_c$ (eq.~\ref{eq:model_N}) and $f_r$ (eq.~\ref{eq:Kmax}) that are generally of order unity. We note that $f_c=4$ in the dynamo case is fairly large, however; this is likely due to the fact that in the MHD dynamo field-line random walk is slightly sub-diffusive (while eq.~\ref{eq:model_N} assumes regular diffusion, which then underpredicts $\langle N \rangle$). Our expression for $N_{\rm m}$ also does not account for trapping between magnetic mirrors. The net-flux cases, for which we find that $f_c\approx 1$ works well, also have trapping in magnetic mirrors, but in these simulations the field-line random walk is super-diffusive due to the mean component (see Section~\ref{sec:net_flux}), which reduces $f_c$.

The good agreement between the model and leaky-box results suggests that the energy dependence of the transport is indeed set by the particles' ability to follow small-scale field-line reversals and the scattering by regions of resonant curvature. Our model and equation~\eqref{eq:model_N} predict a plateau at small $r_{L0}$ set by the random walk of magnetic-field lines. This appears consistent with the small-$r_{L0}$ limit in the MHD-dynamo leaky box, and also with the $\delta B/B_0 \approx 10$ simulation to very good approximation. The well-pronounced turnover in the $\delta B/B_0 \approx 4$ case is a consequence of the finite net flux present in the simulation, which we discuss in more detail in the next section.

\subsection{Impact of a net magnetic flux} \label{sec:net_flux}
In the presence of a net magnetic flux, the transport of particles propagating along tangled field lines is a combination of diffusion and ballistic motion, assuming that the particles constrained by the magnetic fields behave like beads on wires (i.e., if we consider $r_L \rightarrow 0$). In $\delta B/B_0 \gg 1$ turbulence, diffusion is due to the turbulent fluctuations and dominates the transport on small length scales. Ballistic motion is due to the mean component and dominates the transport on large scales, as the expected 3D distance covered by ballistic motion  grows linearly with time while the scaling for diffusion is ${\sim}t^{1/2}$.  

A particle that perfectly follows magnetic-field lines with no pitch-angle diffusion has an effective diffusion coefficient of order $\kappa_{B} \sim c l_B$, where $l_B$ is the coherence length of the field. We can estimate the effective ballistic velocity due to the mean field to be roughly 
\begin{equation}
v_{b} \sim  |\langle c \hat{ \bm b} \rangle | = c |\langle \bb{B} / B\rangle| \sim c |\langle \bb{B} \rangle|/B_{\rm rms},
\end{equation} 
where we have used the fact that the fluctuating component does not contribute to the ballistic velocity (it results in transport better described by diffusion) and we have assumed that the direction of the field and its magnitude are uncorrelated on average. Thus, there is a characteristic length-scale $D$, which separates diffusion-dominated transport from transport dominated by ballistic motion due to the mean field. In particular, $D$ can be estimated by equating the diffusive and ballistic timescales,
\begin{equation}
    \frac{D^2}{\kappa_{B}} \sim \frac{D}{v_b} \ \Rightarrow \ D \sim \kappa_B / {v_b} \sim \frac{B_{\rm rms}}{|\langle \bb{B} \rangle|} l_B,
\end{equation}
and so $D \gg l_B$ in turbulence with $\delta B/B_0 \sim B_{\rm rms}/|\langle \bb{B} \rangle| \gg 1$.

We interpret the turnover in $\langle N \rangle$ in Figure~\ref{fig:leaky_box_meanb} and Figure~\ref{fig:leaky_box_model} as occurring when the CR pitch-angle mean free path exceeds $D$ (and so the problem becomes more similar to the strong-guide-field limit). Consider a pitch-angle mean free path $\lambda_\mu$ evaluated along a magnetic-field line. If $l_B<\lambda_\mu<D$, during one pitch-angle collision time $t_\mu$ the particle covers a three-dimensional distance due to field-line random walk that is $\lambda_{\rm 3D} \sim \sqrt{\kappa_{B} t_\mu} \sim \sqrt{l_B \lambda_\mu }$. Consider now $N_{\rm coll}=\tau/t_\mu $ pitch-angle scattering events during time $\tau$. The expected 3D distance the particle propagates over time $\tau$ is then $\langle d(\tau) \rangle \sim \sqrt{N_{\rm coll}} \lambda_{\rm 3D} \sim \sqrt{\tau / t_\mu} \sqrt{\kappa_{B} t_\mu } \sim \sqrt{\tau \kappa_{B}}$, which is independent of the pitch-angle scattering rate and is set by the field line random walk. If, however, $\lambda_\mu>D$, then $\lambda_{\rm 3D} \sim  v_b t_\mu$. If we again consider $N_{\rm coll}=\tau/t_\mu $ pitch-angle scattering events during time $\tau$, then the expected distance the particle propagates during that time is $\langle d(\tau) \rangle \sim \sqrt{N_{\rm coll}} \lambda_{\rm 3D} \sim \sqrt{\tau / t_\mu}  v_b t_\mu \propto t_\mu^{1/2}$ (which corresponds to a diffusion coefficient $\propto t_\mu$), and so we recover the standard strong guide-field result that pitch-angle diffusion sets the spatial diffusion.

\begin{figure}
  \centering
      \includegraphics[width=0.49\textwidth]{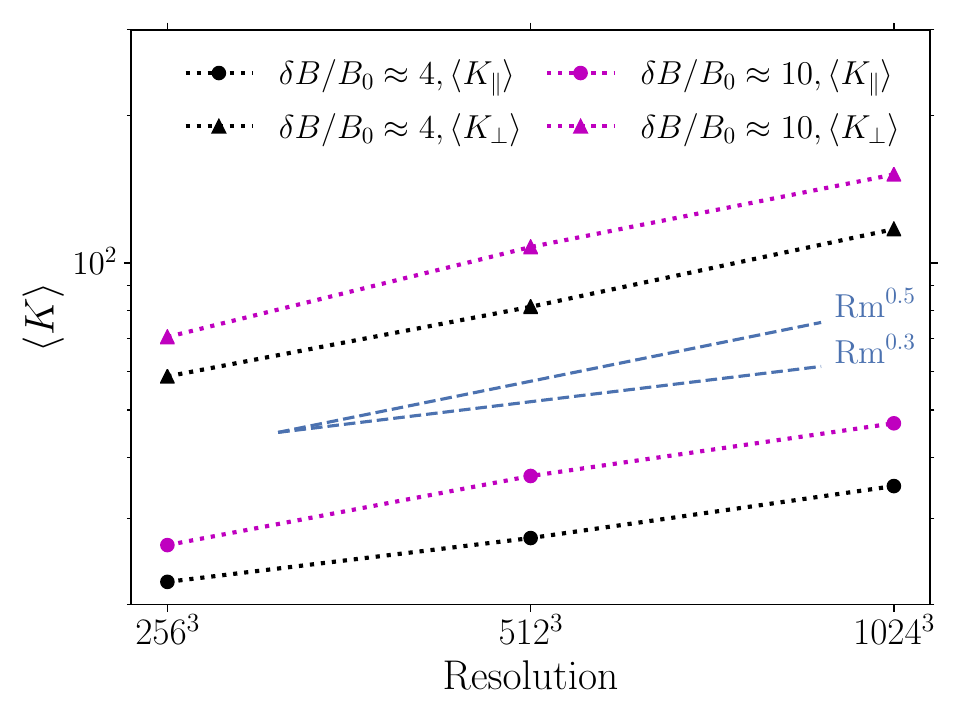}

  \caption{The average field-line curvature $\langle K_\parallel \rangle$ and inverse perpendicular reversal scale $\langle K_\perp \rangle$ in the $\delta B/B_0 \approx 4$ and  $\delta B/B_0 \approx 10$ simulations   at resolutions $256^3$, $512^3$ and $1024^3$. The characteristic field-line reversal/bending scales are not converged with resolution. Also shown is the $\langle K_\perp \rangle \propto {\rm Rm^{0.3}}$ scaling predicted for the tearing-limited dynamo in \citealt{alisa_dynamo}. \label{fig:pdf_resolution}}
\end{figure}

It is very plausible that at higher resolutions the turnover in Figure~\ref{fig:leaky_box_model} occurs at smaller $r_{L0}$. This is partially because, at higher resolutions, the pitch-angle collision time $t_\mu$ at small $r_{L0}$ is likely shorter (as smaller-scale fluctuations are resolved). In addition, at higher resolution, field-line reversals likely occur on smaller scales (see Figure~\ref{fig:pdf_resolution}). This will likely change the energy dependence of $\langle N\rangle$ at small $r_{L0}$ compared to Figure~\ref{fig:leaky_box_model} and expand the dynamic range in particle energy ($r_{L0}$) over which there is significant energy-dependent transport.

\section{Discussion} \label{sec:discussion}
Our results suggest that CR transport in large-amplitude turbulence is fundamentally different from the standard picture of slow pitch-angle diffusion in turbulence with a strong guide field. How important this new physics is for explaining CR observations depends on the volume-filling fraction of $\delta B/B_0 \gg 1$ turbulence, which is currently not very well constrained. While $\delta B/B_0\gg1$ seems unlikely on average in the galactic disc, which has a strong mean component, $\delta B/B_0 \sim$ a few is plausible (\citealt{Haverkorn2015}) and is likely present in at least parts of the Galaxy and its circum-galactic medium. Our findings may thus have significant implications for the transport of CRs in the Milky Way. It is also important to note that the small-scale field reversals generated in such turbulence have been identified as possible sites for the origin of pulsar scintillation and radio wave scattering that (like CR scattering) are very difficult to explain in standard strong guide field turbulence models  (\citealt{goldreich_sridhar_folds, pen_levin_2014}).

\subsection{Dependence on resolution}
A major theoretical uncertainty of our work concerns the characteristic magnetic-field reversal scale in large-amplitude MHD turbulence in the limit of infinite resolution. In particular, by comparing the $K_\parallel$ and $K_\perp$ PDFs in our mean-field simulations at $256^3$, $512^3$ and $1024^3$ we find that the characteristic reversal/bending scales are not converged. This is shown in Figure \ref{fig:pdf_resolution}. \cite{alisa_dynamo} predict that $K_\perp$ in a tearing-limited MHD dynamo scales as ${\rm Rm^{0.3}}$, where Rm is the magnetic Reynolds number. We show that scaling in Figure \ref{fig:pdf_resolution} and it appears to be roughly consistent with our mean-field simulations (at least at the probed resolutions and crudely assuming ${\rm Rm} \propto \ {\rm Resolution}$).\footnote{Our ILES simulations of turbulence with net magnetic flux do not have explicit viscosity or resistivity (only the dynamo simulation uses explicit dissipation). We therefore use resolution to estimate the effective Reynolds numbers. This is motivated by the result that ILES can be described in terms of effective viscous/resistive terms which scale with resolution: $\chi, \eta \propto \Delta_x^{\alpha_{\chi, \eta}}$ where $\chi$, $\eta$ are the viscosity/resistivity, $\Delta_x$ is the grid spacing and the exponents $\alpha_{\chi, \eta}$ have been found to be slightly larger than 1 (the exact values may be different for viscosity and resistivity, and depend on the details of the numerical scheme used; \citealt{zhou_2014}; \citealt{Grete:2023}). Here, for simplicity, we therefore assume that the effective magnetic Reynolds number's dependence on resolution can be crudely approximated by ${\rm Rm} \propto \eta^{-1} \propto 1/\Delta_x \propto {\rm Resolution}$.} It is uncertain whether this scaling continues at even higher resolution. ${\rm Rm}$ is generally enormous in the ISM, which would imply a very small reversal scale $K_\perp^{-1}$, orders of magnitude smaller than the turbulence injection scale. However,  it is not clear how well-motivated the ${\rm Rm}^{0.3}$ scaling is in the presence of a non-zero net flux.   More work on the properties of turbulence in this regime is required.

The characteristic field-line curvature $\langle K_\parallel \rangle $ in Figure~\ref{fig:pdf_resolution} is not converged at our current resolutions.   This suggests that small-Larmor-radius particles (lower-energy CRs) could be very efficiently scattered by small-scale `resonant curvature' structures.   It is, however, unclear that the trend of increasing $\langle K_\parallel \rangle$ 
would continue at even higher resolutions.  That being said, our simulations also do not fully capture the formation of plasmoids in high $\delta B/B_0$ turbulence, which could provide an additional source of high  $\langle K_\parallel \rangle$ CR scattering in simulations at yet higher resolution.

\subsection{Predicted energy dependence of transport}

Figures \ref{fig:leaky_box_meanb} and \ref{fig:leaky_box_model} show that lower-energy particles are better confined than higher energy particles, which is consistent with observations of CRs in the Milky Way.    However, the energy dependence we find at our current resolutions appears to be generally different from the confinement time $r_L^{-0.5}$ scaling suggested by Milky Way observations. The high-Pm dynamo results in Figure \ref{fig:leaky_box_model} are a remarkable exception to this and show a confinement time energy dependence similar to that observed over a large range of particle energy. 

We can use \eqref{eq:nu} to calculate the $P(K_\perp)$ that would produce the `correct' energy dependence. Assuming that asymptotically $\langle B(K_\parallel)\rangle \propto K_\parallel^{-0.5}$, such that $K_{\rm max} \propto r_{L0}^{-2/3}$, we find that,
\begin{equation}
    P(K_\perp) \propto K_\perp^{-1/4} \ \Rightarrow \ \nu(r_{L0}) \propto r_{L0}^{-0.5}.
\end{equation}

For $P(K_\perp) \propto K_\perp^{-1/4}$, $K_\perp P(K_\perp)$ in Figure \ref{fig:pdfs} would scale as $K_\perp^{3/4}$. For the mean-field simulations, however, the actual PDFs for $K_\perp \lesssim \langle K_\perp \rangle$ (i.e. before the peak of the PDF) are significantly steeper than $K_\perp^{3/4}$. This is likely the reason why our mean-field leaky-box results (Figure \ref{fig:leaky_box_model}) deviate significantly from the $r_{L0}^{-0.5}$ scaling. $K_\perp P(K_\perp)$ for the high-Pm dynamo simulation is more consistent with a $K_\perp^{3/4}$ scaling in the range $10\lesssim K_\perp \lesssim 100$, which we believe is the origin of the $r_{L0}^{-0.5}$ scaling for the dynamo in Figure \ref{fig:leaky_box_model}. 

We stress again that the PDF of $K_\perp$ is important not because it is the large $K_\perp$ structures that directly scatter CRs. Instead, a magnetic fold's $K_\perp$ sets whether CRs will be scattered after they traverse a distance $l_B$ (Figure \ref{fig:fold}), as the fold's perpendicular reversal scale is also the scale on which the field eventually bends. If the field-line curvature at the location where the field bends satisfies $K_\parallel c \sim \Omega(B)$, the particles undergo order-unity changes in their pitch angles from scattering. If $K_\parallel c \ll \Omega(B)$, the particles get `scattered' by following the field as it bends while preserving their pitch angle.

We conclude our speculations regarding the energy dependence of transport, and its relation to $P(K_\perp)$, by noting that it is plausible that the PDFs from Figure \ref{fig:pdfs} would enter a new regime at significantly higher resolutions.  Our $\sim 1000^3$ simulations are just beginning to probe the plasmoid instability (see \citealt{fielding_plasmoid} for a detailed resolution study) and do not capture how different plasmoid size distributions affect the shape of the PDFs. In the MHD dynamo plasmoids are expected to form on very small scales $\sim {\rm Rm}^{-0.3}$, where they may play the role of the outer scale of traditional MHD turbulence (\citealt{schekochihin_mhd}). How this carries over to large-amplitude turbulence with a non-zero net flux is unclear. The energy dependence implied by $P(K_\perp)$ may also be affected by the presence of a finite net magnetic flux (Figure \ref{fig:leaky_box_model} and discussion in Section \ref{sec:net_flux}), which will be studied in more detail in follow-up work.     

Future higher-resolution simulations will be essential for identifying whether the novel picture of particle transport found here in regions with $\delta B/B_0 \gtrsim 1$ is important for understanding CR propagation in the Galaxy. Given that mechanisms for efficient scattering in turbulence with a strong guide field remain elusive, this alternative picture of particle transport mediated by reversing magnetic fields seems quite appealing.

\section{Conclusions} \label{sec:conclusions}
In this work, we have studied the propagation of energetic charged particles in large-amplitude MHD turbulence with $\delta B/B_0 \gg 1$. Our motivation for studying this regime is that cosmic-ray (CR) transport models in turbulence with a strong guide field  are not compatible with observations of CRs in the Milky Way (\citealt{kq2022}; \citealt{hopkins_sc_et_problems}). Our primary goal was to identify novel aspects of CR propagation that are not part of standard treatments that focus on small-amplitude fluctuations and quasi-linear theory. In particular, the large $\delta B/B_0 \gg 1$ simulations we consider here correspond to an asymptotic limit of turbulence that is opposite to the $\delta B /B_0 \lesssim 1$ limit usually considered in the CR transport literature. The turbulence in the ISM of galaxies likely ranges from  $\delta B /B_0 \lesssim 1$ to $\delta B /B_0 \sim$ a few. In the latter case, the turbulence may be qualitatively very different from the strong guide field limit and our $\delta B/B_0 \gg 1$ simulations provide clues as to what regulates CR transport in this regime.   

Our turbulence simulations are summarised in Table~\ref{tab:sims}. The turbulence in our simulations creates highly tangled and intermittent magnetic fields (Figures \ref{fig:jz} and \ref{fig:traj_unmag}) with ubiquitous small-scale reversals and bends. Our simulations do not probe a sufficiently large range of scales to properly resolve the disruption of current sheets via the plasmoid instability (e.g.,~\citealt{alisa_dynamo}), and so how plasmoids and their size distribution affect the physics of CR transport is beyond the scope of this paper. In our simulations, the small-scale reversals and bends in the magnetic field are instead due to the stretching and folding of field lines by the large-amplitude turbulence, as in the MHD dynamo (e.g., \citealt{schekochihin_2004}; \citealt{rincon_dynamo}). We have studied the transport of particles in such structures by integrating particle trajectories in stationary snapshots of the turbulence. To quantify the rate of spatial diffusion, both due to pitch-angle diffusion and due to magnetic-field reversals, we used a setup that we term the turbulent leaky box (Section~\ref{sec:leaky_box_setup}). The turbulent leaky box gives us a direct measure of the effective, volume-averaged particle transport rate and it quantifies the combined effects of propagation along tangled magnetic-field lines, pitch-angle scattering and trapping between magnetic mirrors (magnetic bottles).

We find that particle transport in the limit of a weak guide field is qualitatively very different from the strong-guide-field case. Within the range of spatial scales and particle Larmor radii that we can resolve, we find that low-energy particles are confined more efficiently than high-energy particles (Figure~\ref{fig:leaky_box_meanb}), in line with observations.  Remarkably, however, this is not due to faster pitch-angle isotropization at small particle energies; indeed, the opposite is the case (Figure~\ref{fig:t_iso}).  Transport in the limit of a weak guide field is thus fundamentally different from the standard picture of particle diffusion in the presence of a strong guide field, for which confinement and pitch-angle diffusion are effectively synonymous.   We find that, in the weak-guide-field limit, energy-dependent confinement is determined by the energy-dependent (in)ability to follow reversing magnetic-field lines exactly and by scattering in regions of `resonant curvature' (eq.~\ref{eq:kpar}), i.e., where the field line bends on a scale that is of order the local particle gyro-radius.  

We developed a heuristic model for particle transport in magnetic fields organized into folded structures (Section~\ref{sec:model_heuristic} and Figure~\ref{fig:fold}).  The key ingredients in the model are the statistical properties of the field line curvature $K_\parallel$ and the perpendicular reversal scale $K_\perp$. Figure~\ref{fig:pdfs} shows that the perpendicular reversal scale is typically much smaller than the parallel reversal scale. The heuristic model was compared to the turbulent-leaky-box results in Figure~\ref{fig:leaky_box_model}, with overall very good agreement.  In this model, particles in folded magnetic fields are `scattered' at a rate ${\sim}c / l_B$, where $l_B$ is the parallel coherence length of the folds, if their gyro-radii are sufficiently small (eq.~\ref{eq:nu}). This condition makes the transport energy-dependent. The scattering is a combination of pitch-angle scattering by resonant curvature and `scattering' that comes from particles following folded magnetic fields while preserving their pitch angle. The scattering frequency in our model depends on the PDF of $K_\perp$ (eq.~\ref{eq:kperp}), even though large $K_\perp$ does not necessarily imply significant scattering. Instead, a magnetic fold's $K_\perp$ determines whether CRs are scattered after they traverse a distance $l_B$ (Figure~\ref{fig:fold}), as the fold's perpendicular reversal scale is also the scale on which the field eventually bends. 
 
Our findings suggest that the transport of particles in large-amplitude turbulence characterised by small-scale field reversals is fundamentally different from the standard picture of slow pitch-angle diffusion. This may be important for understanding CR transport in the Galaxy.  It is, however,  somewhat unclear whether the energy dependence of transport in magnetic folds is consistent with the energy dependence measured in the Milky Way (see Section~\ref{sec:discussion}) and where in the Milky Way the assumption of weak guide field or zero net flux is relevant. In follow-up work we will study in more detail what the $P(K_\perp)$ PDF may look like in the limit of higher resolution and how it may be affected by resolved tearing and plasmoids of different sizes. It is likely that plasmoids driven by reconnection in $\delta B/B_0 \gtrsim 1$ turbulence contribute an additional source of small-scale `resonant curvature' that can efficiently scatter CRs, even though they are not captured in our current simulations.  The statistical properties of plasmoids in such turbulence may thus be important for understanding the energy-dependence of CR propagation.  Follow-up work will also further quantify what amplitude of turbulence is necessary to generate the small-scale field reversals that drastically change the qualitative picture of CR transport and how this compares to observational constraints in the Milky Way. 

In work related to that presented here, \cite{lemoine_2023} predicts that small-scale bends in magnetic-field lines may be present in strong MHD turbulence even if the guide field is not weak, i.e. $\delta B/B_0 \lesssim 1$, and may be generic intermittent features that exist on top of volume-filling critically-balanced fluctuations. Thus, even in turbulence with a strong guide field, CR transport may be set by rare order-unity scattering events associated with sharp changes in the magnetic-field direction. By calculating PDFs of field-line curvature using coarse-grained turbulent magnetic fields (where the coarse-graining scale reflects the particle gyro-radius), \cite{lemoine_2023} concludes that the volume filling fractions of small-scale bends are sufficient for efficient scattering of low-energy CRs. However, we note that the overall transport rate is most likely a function of both the volume filling fraction and spatial distribution (or clustering) of bends, and so this prediction should be tested further.   In addition, the simulations that \citet{lemoine_2023} uses to test the theory have no net magnetic field and so are more akin to those presented in this paper than to standard strong guide-field simulations.

\section*{Data availability}
The calculations from this article will be shared on reasonable request to the corresponding author.

\section*{Acknowledgements}
We thank Archie Bott, Iryna Butsky, Benjamin Chandran, Philip Hopkins, Rajsekhar Mohapatra, Patrick Reichherzer, Alexander Schekochihin, Anatoly Spitkovsky, Jonathan Squire, Vladimir Zhdankin, Muni Zhou and Ellen Zweibel for useful discussions, which have significantly improved this paper. We thank the referee for an insightful report that led to an improved presentation. PK was supported by the Lyman Spitzer, Jr. Fellowship at Princeton University. MWK was supported in part by NSF CAREER award No.~1944972. EQ was supported in part by a Simons Investigator grant from the Simons Foundation and by NSF AST grant 2107872. DF's and BR's research at the Flatiron Institute was supported by the Simons Foundation. BR's support for this work was also provided by NASA through the NASA Hubble Fellowship grant HST-HF2-51518.001-A awarded by the Space Telescope Science Institute, which is operated by the Association of Universities for Research in Astronomy, Incorporated, under NASA contract NAS5-26555.  This research was facilitated by the Multimessenger Plasma Physics Center (MPPC), NSF grants PHY-2206607 and PHY-2206610. The net-flux simulations used in this work were performed at facilities supported by the Scientific Computing Core at the Flatiron Institute, a division of the Simons Foundation. The dynamo simulation taken from \citet{alisa_dynamo} was performed as part of the Frontera computing project at the Texas Advanced Computing Center. We also made extensive use of the Stellar cluster at the PICSciE-OIT TIGRESS High Performance Computing Center and Visualization Laboratory at Princeton University.

\bibliographystyle{mnras}

\bibliography{test_particle}



\bsp	
\label{lastpage}
\end{document}